\documentclass[pra,superscriptaddress,twocolumn,showpacs,preprintnumbers,amsmath,amssymb]{revtex4}

\usepackage[dvips]{graphicx}
\usepackage{dcolumn}
\usepackage[normalem]{ulem}

\usepackage{amsmath}
\usepackage{amssymb}
\usepackage{upgreek}
\usepackage{psfrag}

\newcommand{\bm}[1]{\boldsymbol{\mathbf{#1}}}
\newcommand{\ud}{\mathrm{d}}
\newcommand{\bra}{\left\langle}
\newcommand{\ket}{\right\rangle}
\newcommand{\vbar}{\bar{v}}

\newcommand{\NoAutoSpaceBeforeFDP}[0]{}
\newcommand{\AutoSpaceBeforeFDP}[0]{}

\begin{document}

\preprint{}

\title{Enhancement of radiation trapping for quasi-resonant scatterers at low temperature}

\author{Romain Pierrat}
\affiliation{Laboratoire Kastler-Brossel, Universit\'e Pierre et Marie Curie, ENS, CNRS;
4 Place Jussieu, F-75005 Paris, France}
\email{romain.pierrat@spectro.jussieu.fr}
\author{Beno\^it Gr\'emaud}
\affiliation{Laboratoire Kastler-Brossel, Universit\'e Pierre et Marie Curie, ENS, CNRS;
4 Place Jussieu, F-75005 Paris, France}
\affiliation{IPAL, CNRS; 1 Fusionopolis Way, Singapore 138632, Singapore}
\affiliation{Centre for Quantum Technologies, National University of Singapore, 3 Science Drive 2, Singapore}
\author{Dominique Delande}
\affiliation{Laboratoire Kastler-Brossel, Universit\'e Pierre et Marie Curie, ENS, CNRS;
4 Place Jussieu, F-75005 Paris, France}

\date{\today}

\begin{abstract}
   We present a transport equation for the incoherent propagation of radiation inside
   a quasi-resonant atomic gas at low temperature. The derivation is based on a
   generalized Bethe-Salpeter equation taking into account the motion of the 
   atoms. The obtained equation is similar to the radiative transfer equation. 
   It is solved numerically by an original Monte Carlo approach in the case of a slab geometry. The 
   partial frequency redistribution caused by the small velocity of the scatterers
   make the emitted flux outside the system and 
   the energy density inside the medium to behave differently than in the case
   of complete frequency redistribution. In particular, the long
   time dependence of the specific intensity (escape factor) is slightly different from the Holstein prediction.
\end{abstract}

\pacs{42.25.Dd, 32.80.-t}

\maketitle

\section{Introduction}

The study of light propagation in a scattering system has become a very active field of research
in mesoscopic physics for few years because of its many applications in particular in biomedical
imaging~\cite{SEBBAH-2001}. Lots of key features like Anderson localization of photons~\cite{ANDERSON-1958}
or fluorescence lifetime of a single emitter embedded in a complex system have to be understood
in such a system~\cite{FROUFE-2007}. A very good medium to deal with these fundamental phenomena is composed of
cold atoms. Cold atoms can be manipulated with a high degree of control in particular to realize
a monodisperse ensemble of strongly resonant point scatterers, free of defects and absorption.

Since the thirties and the pioneering work of Kenty~\cite{KENTY-1932}, the trapping of
photons in a scattering and resonant system is well understood. For a dilute system
illuminated by a laser beam and in the limit of
low saturation of the atomic transition (e.g. at low light intensity for resonant light) and sufficiently
low density for atom-atom collisional broadening to be neglected,
the scattering remains fully elastic. Thus, during its
propagation through the system, the light remains monochromatic and the scattering mean-free
path can be continuously adjusted by tuning the laser frequency. It is then possible to observe
coherent effect such as weak and strong localization~\cite{KAISER-1999}.

At very low temperature and if the system is not too large, the residual motion of the atoms does
not modify significantly the frequency of a scattered photon and coherent effects such 
as coherent backscattering~\cite{KAISER-1999,KUPRIYANOV-2003}
can still be observed, as if the atoms were pinned at fixed positions. 
Residual motion
even provides us with a significant advantage: averaging over different disorder realizations is done
automatically by just waiting for the atoms to move by say one wavelength, i.e. by trivial time averaging.
 
The Doppler shift of the scattered photon can be neglected if the frequency change does not modify the propagation
properties, meaning that the scattered photon can be rescattered like the incoming ones.
The scattering cross-section varies rapidly around an atomic resonance at frequency $\omega_0$ over a frequency range
$\Gamma,$ where $\Gamma$ is the inverse of the lifetime of the atomic excited state. Thus, if $k$ denotes
the wave factor of the incoming photons and $\vbar$ the typical atomic velocity, the typicla Doppler shift
after a scatteting event is $k\vbar$ and 
the regime of ``quasi-pinned'' disorder
is reached when:
\begin{equation}
 k\vbar \ll \Gamma
\label{validity-naive}
\end{equation}
 As will be seen in the following, this is only a very rough criterion and a more careful analysis
is needed.
Eq.~(\ref{validity-naive}) can be rewritten as $\vbar/c \ll \omega_0/\Gamma,$ the latter quantity
being the quality factor of the atomic resonance, a number typically of the order of $10^8.$ 
This thus puts a severe limit on the atomic velocity, of the order of 1 m/s, and explains
why observation of coherent effects requires cold atomic gases. 

However, even if the temperature is low, for a sufficiently large medium,
many scattering events can be chained before a photon escapes. Even if the
Doppler shift of a single scattering event is small, the global frequency shift
accumulated along a multiple scattering path can be non negligible.
This depends of course on the optical thickness $b$:
\begin{equation}
 b =\frac{L}{\ell}
\end{equation}
where $L$ is the size of the medium and $\ell$ the scattering mean free path of the photon
in the medium.
At large $b$, a typical multiple scattering path can be viewed as a random walk of the photon
in the medium with step $\ell.$ Thus the photon is multiply scattered about $b^2$ times before
escaping. As the Doppler shift depends on the relative orientation of the atomic velocity and the incoming and outgoing
wave vectors of the photon, it is on average a random variable with zero average and 
about $(k\vbar)^2$ variance. If successive scattering events are statistically independent,
the photon frequency performs itself a random walk with step about $k\vbar,$
and the typical accumulated Doppler shift after $N$ scattering events is of the order
of $\sqrt{N}k\vbar.$ As $N\sim b^2,$ a rough criterion for neglecting
this effect becomes:
\begin{equation}
 b k \vbar \ll \Gamma.
\label{validity}
\end{equation}

Some
theoretical works has been done previously in particular in rather dense atomic vapors
typical of interstellar atmospheres, when collisional
broadening is larger than the natural linewidth of the atomic resonance. In such a regime,
the scattering process is incoherent and inelastic: the frequency of the scattered photon
is essentially decorrelated from the incoming frequency. This is the complete frequency redistribution (CFR) 
regime~\cite{HOLSTEIN-1947,HOLSTEIN-1951,MOLISCH-1998}. Depending on parameters
such as the optical thickness and the Doppler broadening, several regimes can 
be obtained, leading to various decay rate
equations such as the Holstein's equation~\cite{PAYNE-1974}. 
In that case, the photon frequency can be Doppler shifted
very far from resonance, so that the medium becomes almost transparent:
this is a regime where photons trajectories can be L\'evy flights~\cite{PEREIRA-2004}.

The present work deals with the case of low temperature (or slow atoms) for which we 
cannot assume a complete frequency redistribution.
The partial frequency redistribution (PFR) is then due to Doppler effect. This is an interesting new regime 
where the radiation transport has both coherent (the individual scattering event)
and incoherent (because of randomization due to the atomic motion) aspects. 
Moreover, the study of such a system is of primary importance
to design new high capacity quantum memories using cold atoms, in particular to give a reliable expression
of the escape factor (i.e. exponential time decay rate at long times).
The main idea is to derive a transport
equation for the incoherent radiation from first principles in the case of moving atoms,
taking into account the resonant character of the scatterers. This is
very different from e.g. diffusing-wave
spectroscopy theory~\cite{PINE-1990,ACKERSON-1992} or electron transport~\cite{AKKERMANS-2007}.
It is also expected that the dynamics of the atoms affect the coherent properties of the system as shown
experimentally~\cite{LABEYRIE-2006}. Theoritical and numerical works are in progress in particular to describe the thermal decoherence
of the backscattering cone.

More precisely, we deal with systems composed of a cloud of two-level atoms: this is a very good
approximation for quasi-resonant atomic scatterers as e.g. Rubidium atoms used in experiments.
As we consider here only the incoherent transport of radiation, the existence of an internal
hyperfine structure which leads to considerable modifications of the coherent
transport such as coherent back scattering~\cite{MUELLER-2001}, is irrelevant.
We assume a dilute gas where the scattering mean free path is much larger than the optical wavelength.
We will also treat atoms as classical point scatterers. This assumption
certainly fails at very low temperature where the atomic de Broglie
wavelength becomes comparable to the optical wavelength: in this regime,
the recoil effect induced by the scattering of a single photon
significantly affects the motion of the center of mass of the atom.
We thus exclude ultra-cold atomic gases. Properly taking into account the quantum nature
of the external atomic motion is a much more complicated problem, 
see e.g.~\cite{WICKLES-2006} for the simple case of two atoms.
For usual atoms like Rubidium, the recoil velocity is of the order of few mm/s, and
one can easily have cold atoms faster than the recoil velocity, but still obeying
Eq.~(\ref{validity-naive}).
We also exclude collective quantum effects such as Bose-Einstein condensation, taking place when the
atomic de Broglie wavelength is comparable to the inter-atomic distance.

We are interested in typically an atomic cloud in a magneto-optical trap (MOT), scattering photons
from a laser beam. We will thus assume a thermal distribution of the atomic velocities. 
In such a system, the radiation felt deep in the medium is quasi-isotropic such that atoms are not
significantly pushed by incoming photons. On the other handm, in a real experiment, 
if the laser beam is sufficiently
intense, the accumulated recoil effect may significantly disturb the atomic velocity
distribution. For simplicity, we neglect this effect, although taking it into account would be easy in
the present framework.

The paper is organized as follows:
in Sect.~\ref{first_principles}, we start from first principles to
obtain a generalized form of the well-known Dyson (for the average field) and Bethe-Salpeter 
(for the field correlations, including the average intensity) equations. 
Then, some assumptions are done in Sect.~\ref{assumptions}
to derive the transport equation which is the main result of this paper. This equation is solved numerically
in Sect.~\ref{numerical_sim} using an original Monte Carlo scheme. Finally, Sect.~\ref{modal_approach} is devoted
to the modal study of the transport equation to explicitly derive the spectral and temporal behaviors at large scales.
In particular, we show that, although the spatial motion of the photons is in general not diffusive, 
it is possible to derive equations of the Fokker-Planck type, which govern
the evolution of the intensity distribution as a function of position and frequency.
This section ends up with the simple picture story of photons migration valid at large scales.

\section{Dyson and Bethe-Salpeter equations for moving scatterers}\label{first_principles}

\subsection{Scattering operator}

Starting with first principles means that we have to derive a generalized form of the scattering operator taking
into account the Doppler effect. It
will be the fundamental quantity of our derivation. As mentioned above,
the recoil is not taken into account in this derivation because
the recoil-induced drift of the frequency distribution is usually much smaller than the Doppler-induced spreading
($v_{\textrm{recoil}}/\vbar \sim 10^{-3}-10^{-2}$). 
Note that going beyond this approximation is far from trivial.
We denote by $t\left(\bm{k},\bm{k}',\omega\right)$ the scattering operator
for a single
fixed atom. $\bm{k}$ and $\bm{k}'$ are the incident and scattered wave-vectors respectively and $\omega$ is the
frequency. The reader will find in Appendix~\ref{fourier_transforms} the conventions used for the
spatio-temporal Fourier transforms of all quantities. For the sake of simplicity,
we assume that all the scattering atoms are identical. Let us consider an atom moving at velocity
$\bm{v}$ constant during the whole scattering process. 
We denote by $t_m^{\bm{v}}\left(\bm{k},\bm{k}',\omega,\omega'\right)$ 
the scattering operator for this moving atom (subscript $m$ means \emph{moving}). 
Thus, the incident ($E_{\textrm{inc}}$) and scattered ($E_{\textrm{sca}}$) fields are related by
\begin{align}\nonumber
   E_{\textrm{sca}}\left(\bm{k},\omega\right)=\int & G_0\left(\bm{k},\omega\right)
   t_m^{\bm{v}}\left(\bm{k},\bm{k}',\omega,\omega'\right)
\\\nonumber & \times
   E_{\textrm{inc}}\left(\bm{k}',\omega'\right)
   \frac{\ud^3\bm{k}'}{8\pi^3}\frac{\ud\omega'}{2\pi}
\end{align}
in the scalar approximation (polarization effects neglected), 
where $G_0$ is the Green function in the vacuum. 
Note that taking into account polarization effects is a straightforward 
extension of the present analysis. Indeed, Doppler effect acts similarly on all polarization
components. It is thus sufficient to add the angular dependence of the scattering cross-section 
in all formulas. We drop this dependence for simplicity and focus on the
main features related to the motion of the scatterers.

Considering now this expression in the atom frame, we have 
\begin{align}\label{scattering}
   & \mathcal{E}_{\textrm{sca}}\left(\bm{k},\omega\right)=\int G_0\left(\bm{k},\omega+\bm{k}\cdot\bm{v}\right)
\\\nonumber & \hspace{1cm}\times
   t_m^{\bm{v}}\left(\bm{k},\bm{k}',\omega+\bm{k}\cdot\bm{v},\omega'+\bm{k}'\cdot\bm{v}\right)
\\\nonumber & \hspace{1cm}\times
   \mathcal{E}_{\textrm{inc}}\left(\bm{k}',\omega'\right)
   \frac{\ud^3\bm{k}'}{8\pi^3}\frac{\ud\omega'}{2\pi}
\end{align}
where $\mathcal{E}$ is the field in the atom frame in which the atom is fixed. 
The relation between the incident and the scattered fields can be written in the form
\begin{equation}\label{scattering2}
   \mathcal{E}_{\textrm{sca}}\left(\bm{k},\omega\right)= \int G_0\left(\bm{k},\omega\right)
   t\left(\bm{k},\bm{k}',\omega\right)
   \mathcal{E}_{\textrm{inc}}\left(\bm{k}',\omega\right)
   \frac{\ud^3\bm{k}'}{8\pi^3}.
\end{equation}
By identifying equations~(\ref{scattering}) and~(\ref{scattering2}) 
and taking into account that the velocity of the atom is weak enough 
to have $\bm{k}\cdot\bm{v}\ll\omega$ in the Green operator, we have 
\begin{align}\nonumber
   t_m^{\bm{v}}\left(\bm{k},\bm{k}',\omega,\omega'\right)
   = & 2\pi t\left(\bm{k},\bm{k}',\omega-\bm{k}\cdot\bm{v}\right)
\\\label{scattering3} & \times
   \bm{\delta}\left[\omega'-\omega-\left(\bm{k}'-\bm{k}\right)\cdot\bm{v}\right],
\end{align}
which is the expression of the generalized scattering operator, a fundamental quantity in our derivation.
$\bm{\delta}$ denotes the Dirac-delta function.
The physical interpretation of $t_m$ is straightforward: the modification of frequency in the $t$ operator
corresponds to the Doppler shifted frequency in the atomic frame, while the $\bm{\delta}$
function expresses that the scattering is elastic in the atomic frame. The latter expresses a strong
correlation between the incoming and the outgoing frequency, in strong contrast with the usual
complete frequency redistribution hypothesis~\cite{HOLSTEIN-1947,HOLSTEIN-1951}. 

\subsection{Two-levels atom polarizability}

The scalar scattering operator for pinned point scatterers is related to the polarizability by
\begin{equation}\nonumber
   t\left(\bm{k},\bm{k}',\omega\right)=-\frac{\omega^2}{c_0^2}\alpha\left(\omega\right)
\end{equation}
where $c_0$ is the light velocity in vacuum.
The system is composed of two-level atoms with resonant frequency $\omega_0$. $\Gamma$
 is the spontaneous emission decay rate. The incoming photons are quasi-resonant,
 such that $\delta=\omega-\omega_0\ll\omega_0$. The two-level atom
scalar polarizability is then given by
\begin{equation}\label{polar}
   \alpha\left(\omega\right)=-\frac{4\pi}{k_0^3}\frac{\Gamma/2}{\delta+i\Gamma/2}
\end{equation}
where $k_0=\omega_0/c_0$ is the wave-vector in vacuum at frequency $\omega_0$. We thus have
\begin{equation}
\label{tvm}
   t\left(\bm{k},\bm{k}',\omega\right)
   \sim\frac{4\pi}{k_0}\frac{\Gamma/2}{\delta+i\Gamma/2}.
\end{equation}
Thus, the generalized scattering operator is obtained by plugging Eq.~(\ref{tvm}) in Eq.~(\ref{scattering3}).
The main differences between this equation and the expression of the scattering operator for fixed atoms
are the Dirac $\delta$ function representing the Doppler frequency shift.

\subsection{Dyson equation}

The Dyson equation is a closed equation for the field $\bra E\ket$ averaged over an ensemble of realization of
the scattering medium~\cite{DYSON-1949a,DYSON-1949b}. For the case of fixed atoms,
the average is performed over the positions of the scatterers.
For moving atoms, the average is performed over the initial positions, times and velocities
of the scatterers. In this case, the generalized form of the Dyson equation is given by
\begin{align}\nonumber
   \bra E\ket\left(\bm{r},t\right)=E_{\textrm{inc}}\left(\bm{r},t\right)
   +\int & G_0\left(\bm{r}-\bm{r}',t-t'\right) M_m\left(\bm{r}',\bm{r}'',t',t''\right)
\\ & \times
   \bra E\ket\left(\bm{r}'',t''\right)
   \ud^3\bm{r}'\ud^3\bm{r}''\ud t'\ud t'',
\end{align}
where $M_m$ is called the mass operator. 
It essentially describes the extinction phenomenon due
to scattering. In particular, it
contains the expression of the scattering mean-free path depicting the exponential decay of the averaged (or
coherent) field. The mass operator is the equivalent of the self-energy for the Schr\"odinger equation
for matter waves in a random potential~\cite{AKKERMANS-2007}.
For fixed scatterers, one usually considers a statistically translationally
invariant infinite medium.
In our case, we also assume that the atom velocity distribution is time-independent
and uncorrelated with the position, so that
\begin{equation}\nonumber
   M_m\left(\bm{r}',\bm{r}'',t',t''\right)=M_m\left(\bm{r}'-\bm{r}'',t'-t''\right).
\end{equation}
This expression allows us to rewrite the Dyson equation in term of a Fourier transform:
\begin{equation}\label{dyson}
      \bra E\ket\left(\bm{k},\omega\right)=E_{\textrm{inc}}\left(\bm{k},\omega\right)
      +G_0\left(\bm{k},\omega\right)M_m\left(\bm{k},\omega\right)\bra E\ket\left(\bm{k},\omega\right).
\end{equation}
Writing the same equation for the averaged Green function, we obtain
\begin{equation}\nonumber
      \bra G\ket\left(\bm{k},\omega\right)=G_0\left(\bm{k},\omega\right)
      +G_0\left(\bm{k},\omega\right)M_m\left(\bm{k},\omega\right)\bra G\ket\left(\bm{k},\omega\right)
\end{equation}
which gives the expression
\begin{equation}\label{averaged_green}
   \bra G\ket\left(\bm{k},\omega\right)=\frac{1}{\omega^2/c_0^2-k^2-M_m\left(\bm{k},\omega\right)}
\end{equation}
considering that $G_0\left(\bm{k},\omega\right)=1/\left(\omega^2/c_0^2-k^2\right)$.
Then we can define a wave-vector in the effective homogeneous medium
$k_{\textrm{eff}}^2=\omega^2/c_0^2-M_m\left(\bm{k_{\textrm{eff}}},\omega\right)$, 
the real and imaginary part of which give the effective wavelength and the scattering mean-free path respectively.
To derive the expression of the mass operator we assume that the system is dilute enough for the first order
diagrammatic expansion of the operator to be valid. This is the Foldy-Twersky approximation (or the Born
approximation)~\cite{FOLDY-1945}. In that case, the
mass operator is the sum over all the $N$ scatterers of the averaged scattering operator for all accessible
positions, times and velocities:
\begin{align}\nonumber
   M_m\left(\bm{r}-\bm{r}',t-t'\right)=\sum_{i=1}^N\int &
   t_m^{\bm{v}_i}\left(\bm{r}-\bm{r}_i,\bm{r}'-\bm{r}_i,t-t_i,t'-t_i\right)
\\ & \times
   P\left(\bm{r}_i,t_i,\bm{v}_i\right)
   \ud^3\bm{r}_i\ud t_i\ud^3\bm{v}_i
\end{align}
where $P\left(\bm{r},t,\bm{v}\right)$ is the probability density to have a scatterer of velocity $\bm{v}$ at
position $\bm{r}$ and at time $t$. 
For our translationally invariant medium,
$P\left(\bm{r},t,\bm{v}\right)=g\left(\bm{v}\right)/\left(VT\right)$ where $V$ is the volume of the system and
$T$ the time window. In the large $V,T$ limit, the mass operator in the Fourier domain reads
\begin{equation}\nonumber
   M_m\left(\bm{k},\omega\right)=\lim_{T\to\infty}\frac{\rho}{T}\int
   t_m^{\bm{v}}\left(\bm{k},\bm{k},\omega,\omega\right)g\left(\bm{v}\right)\ud\bm{v}
\end{equation}
where $\rho=N/V$ is the density of scatterers. This leads to the final expression of the mass 
operator if we remark that $\lim_{T\to\infty}2\pi\bm{\delta}\left(\omega-\omega\right)/T=1$:
\begin{equation}\label{mass}
      M_m\left(\bm{k},\omega\right) = \rho \int t\left(\bm{k},\bm{k},\omega-\bm{k}\cdot\bm{v}\right)
      g\left(\bm{v}\right)\ud\bm{v}.
\end{equation}
For a thermal distribution, this gives
a standard Voigt profile for absorption~\cite{voigt}, which itself reduces
to a usual Doppler Gaussian profile in 
the limit $k\vbar \gg \Gamma.$ In the latter case, the averaging over the atomic velocity
makes the correlation between the incoming and scattered
frequencies much smaller than for slow atoms.

\subsection{Bethe-salpeter equation}

The Bethe-Salpeter equation is a closed equation for the field autocorrelation function 
$\bra EE^*\ket$~\cite{RYTOV-1989,APRESYAN-1996,FRISCH-1968}. 
It contains an operator $K$ depending on four space variables called 
the intensity (or vertex) operator and describing the correlation between two scattering processes.
As for the Dyson equation, the idea is to generalize this equation to the case of moving scatterers. This leads
to a new vertex operator $K_m$ depending on four space variables and four time variables. We assume that no
source is present (the incident current densities associated with the coherent term are missing).
Thus the generalized Bethe-Salpeter equation writes:

\begin{widetext}
   \begin{align}\nonumber
         \bra E\left(\bm{r}_1',t_1'\right)E^*\left(\bm{r}_2',t_2'\right)\ket= &
         \int \bra G\left(\bm{r}_1',\bm{r}_1,t_1',t_1\right)\ket\bra G^*\left(\bm{r}_2',\bm{r}_2,t_2',t_2\right)\ket
         K_m\left(\bm{r}_1,\bm{r}_3,\bm{r}_2,\bm{r}_4,t_1,t_3,t_2,t_4\right)
   \\ & \hphantom{\int}\times
         \bra E\left(\bm{r}_3,t_3\right)E^*\left(\bm{r}_4,t_4\right)\ket
         \ud^3\bm{r}_1\ud^3\bm{r}_2\ud^3\bm{r}_3\ud^3\bm{r}_4
         \,\ud t_1\ud t_2\ud t_3\ud t_4,
   \end{align}
   where the symbol $^*$ denotes the conjugate quantity. The global translational invariance in space and time
   for the averaged quantities leads to a factorization of the vertex operator as follows
   \begin{align}\nonumber
      K_m\left(\bm{k}_1,\bm{k}_3,\bm{k}_2,\bm{k}_4,\omega_1,\omega_3,\omega_2,\omega_4\right)= &
      8\pi^3\bm{\delta}\left(\bm{k}_1-\bm{k}_3-\bm{k}_2+\bm{k}_4\right)
      \times 2\pi\bm{\delta}\left(\omega_1-\omega_3-\omega_2+\omega_4\right)
   \\ &
      \times\widetilde{K}_m\left(\bm{k}_1,\bm{k}_3,\bm{k}_2,\bm{k}_4,\omega_1,\omega_3,\omega_2,\omega_4\right).
   \end{align}
   This leads to the following expression of the Fourier transform of the Bethe-Salpeter equation for an infinite medium:
   \begin{align}\nonumber
      f\left(\bm{k},\bm{q},\omega, \Omega\right) = & 
      \int\bra G\left(\bm{k}+\frac{\bm{q}}{2},\omega+\frac{\Omega}{2}\right)\ket
      \bra G^*\left(\bm{k}-\frac{\bm{q}}{2},\omega-\frac{\Omega}{2}\right)\ket
   \\\label{bethe-salpeter} &\hphantom{\int}
      \times\widetilde{K}_m\left(\bm{k}+\frac{\bm{q}}{2},
      \bm{k}'+\frac{\bm{q}}{2},\bm{k}-\frac{\bm{q}}{2},\bm{k}'-\frac{\bm{q}}{2},
      \omega+\frac{\Omega}{2},\omega'+\frac{\Omega}{2},\omega-\frac{\Omega}{2},\omega'-\frac{\Omega}{2}\right)
      f\left(\bm{k}',\bm{q},\omega',\Omega\right)\frac{\ud^3\bm{k}'}{8\pi^3}\frac{\ud\omega'}{2\pi}
   \end{align}
   where the function $f$ is the Fourier form of the spatio-temporal correlation function of the electric
   field (i.e. $f\left(\bm{k},\bm{q},,\omega, \Omega\right)=
   \bra E\left(\bm{k}+\bm{q}/2,\omega+\Omega/2\right)E^*\left(\bm{k}-\bm{q}/2,\omega-\Omega/2\right)\ket$).

   As for the mass operator, we only keep the first term of the diagrammatic expansion of the vertex operator
   (Ladder approximation valid for a dilute gas) which leads to a sum over all scatterers,
   and an averaging over all accessible positions, times and velocities of the scattering operator correlation function. 
   In Fourier space, we obtain:
   \begin{equation}\nonumber
      \widetilde{K}_m\left(\bm{k}_1,\bm{k}_3,\bm{k}_2,\bm{k}_4,\omega_1,\omega_3,\omega_2,\omega_4\right)=
      \lim_{T\to\infty}\frac{\rho}{T}\int
      t_m^{\bm{v}}\left(\bm{k}_1,\bm{k}_3,\omega_1,\omega_3\right)
      t_m^{\bm{v}*}\left(\bm{k}_2,\bm{k}_4,\omega_2,\omega_4\right)
       g\left(\bm{v}\right)\ud\bm{v}.
   \end{equation}
   The averaging over the atomic velocity is performed over the product $t_m^{\bm{v}}t_m^{\bm{v}*}$ 
   to take into account the fact that
   the scattering process occurs on the same atom for the field and its conjugate.
   Using this expression of the scattering operator in Eq.~(\ref{scattering3}), the intensity operator becomes
   \begin{align}\nonumber
      & \widetilde{K}_m\left(\bm{k}+\frac{\bm{q}}{2},\bm{k}'+\frac{\bm{q}}{2},
      \bm{k}-\frac{\bm{q}}{2},\bm{k}'-\frac{\bm{q}}{2},\omega+\frac{\Omega}{2},
      \omega'+\frac{\Omega}{2},\omega-\frac{\Omega}{2},\omega'-\frac{\Omega}{2}\right)
      \rho\int t\left[\bm{k}+\frac{\bm{q}}{2},\bm{k}'+\frac{\bm{q}}{2},\omega+\frac{\Omega}{2}-\left(\bm{k}+\frac{\bm{q}}{2}\right)\cdot\bm{v}\right]
   \\\label{vertex} & \hspace{1cm}\hphantom{\rho\int}\times
             t^*\left[\bm{k}-\frac{\bm{q}}{2},\bm{k}'-\frac{\bm{q}}{2},\omega-\frac{\Omega}{2}-\left(\bm{k}-\frac{\bm{q}}{2}\right)\cdot\bm{v}\right]
      2\pi\bm{\delta}\left[\omega'-\omega-\left(\bm{k}'-\bm{k}\right)\cdot\bm{v}\right]g\left(\bm{v}\right)\ud\bm{v}.
   \end{align}

   \section{Transport equation}\label{assumptions}

   The transport equation we will obtain is an equation governing the specific intensity 
   $I\left(\bm{r},\bm{u},t,\omega\right)$
   inside the system. This is a local ($\bm{r}$) and directional ($\bm{u}$) radiative flux at 
   time $t$ and frequency $\omega$.
   This quantity will be defined using the field autocorrelation function. So the root of the 
   derivation is the Bethe-Salpeter
   equation in the form obtained in Eq.~(\ref{bethe-salpeter}). 
   To exhibit the spatio-temporal derivatives of the specific
   intensity, we can transform the product of the averaged Green functions in a 
   difference: $1/(AB)=(1/A-1/B)/(B-A),$ and use the averaged Green function in Eq.~(\ref{averaged_green}).
   We obtain:
   \begin{align}\label{green}
      \bra G\left(\bm{k}+\frac{\bm{q}}{2},\omega+\frac{\Omega}{2}\right)\ket
      \bra G^*\left(\bm{k}-\frac{\bm{q}}{2},\omega-\frac{\Omega}{2}\right)\ket
      =
      \frac{\displaystyle \bra G\left(\bm{k}+\frac{\bm{q}}{2},\omega+\frac{\Omega}{2}\right)\ket-
      \bra G^*\left(\bm{k}-\frac{\bm{q}}{2},\omega-\frac{\Omega}{2}\right)\ket}
      {\displaystyle -\frac{2\Omega\omega}{c_0^2}+2\bm{k}\cdot\bm{q}+
      M_m\left(\bm{k}+\frac{\bm{q}}{2},\omega+\frac{\Omega}{2}\right)-
      M_m^*\left(\bm{k}-\frac{\bm{q}}{2},\omega-\frac{\Omega}{2}\right)}.
   \end{align}
\end{widetext}

The effective wave-vector in the medium is given by
$k_{\textrm{eff}}^2=\omega^2/c_0^2-M_m\left(\bm{k_{\textrm{eff}}},\omega\right)$. By replacing the mass operator by its
approximate expression, Eq.~(\ref{mass}) and the polarizability by Eq.~(\ref{polar}), we obtain
\begin{align}\nonumber
   \Re\left[k_{\textrm{eff}}^2\right] & =\frac{\omega^2}{c_0^2}-\frac{4\pi\rho}{k_0}
   \int\frac{\Gamma/2\left(\delta-\bm{k}\cdot\bm{v}\right)}{\left(\delta-\bm{k}\cdot\bm{v}\right)^2+\Gamma^2/4}g\left(\bm{v}\right)\ud\bm{v},
\\\nonumber
   \Im\left[k_{\textrm{eff}}^2\right] & =\frac{4\pi\rho}{k_0}
   \int\frac{\Gamma^2/4}{\left(\delta-\bm{k}\cdot\bm{v}\right)^2+\Gamma^2/4}g\left(\bm{v}\right)\ud\bm{v}.
\end{align}
In a dilute medium such that
$\rho\lambda_0^3\ll 1$ where $\lambda_0=2\pi/k_0$ is the wavelength in the vacuum at frequency $\omega_0$, we have
$\Im\left[k_{\textrm{eff}}^2\right]\ll\Re\left[k_{\textrm{eff}}^2\right]\sim \omega^2/c_0^2$. Writing 
$k_{\textrm{eff}}=k'+ik''$ we find
\begin{align}
   k' & =\sqrt{\Re\left[k_{\textrm{eff}}^2\right]}=\frac{2\pi}{\lambda_{\textrm{eff}}}\sim\frac{\omega}{c_0},
\\\nonumber
   k'' & =\frac{\Im\left[k_{\textrm{eff}}^2\right]}{2\sqrt{\Re\left[k_{\textrm{eff}}^2\right]}}=\frac{1}{2\ell}\ll k'
\end{align}
where $\lambda_{\textrm{eff}}$ is the effective wavelength (close to the vacuum wavelength
in a dilute medium) and $\ell\gg \lambda_0$ the scattering mean-free path.
Physically,
this means that the spatial variations of the specific intensity take place on a scale
much longer than the wavelength. We can thus restrict ourselves to $q\ll k,k'$ in Eq.~(\ref{bethe-salpeter}).
Similarly, the temporal variations of
the specific intensity are slow (typically of the order of $\Gamma^{-1}$) compared to the 
temporal variations of the wave (on a time scale $\omega_0^{-1})$.
This allows to define properly the specific intensity with two independent spatial variables (position
$\bm{r}$ and direction $\bm{u}$) and two independent temporal variables (time $t$ and frequency $\omega$)
and assume $\Omega\ll\omega$ in Eq.~(\ref{bethe-salpeter}). Note however that we do NOT assume
$\Omega\ll \Gamma$ --- meaning that we keep the full dynamics on the time scale
$\Gamma^{-1}$ --- and that because the polarizability
varies around the resonance on a scale $\Gamma$, we must keep explicitly this dependence.
Using these approximations coupled to Eqs.~(\ref{mass}), (\ref{vertex}) and
(\ref{green}), the Bethe-Salpeter equation~(\ref{bethe-salpeter}) reduces to
\begin{widetext}
   \begin{align}\nonumber
      & \left[-\frac{2\Omega\omega}{c_0^2}+2\bm{k}\cdot\bm{q}-\rho\int\left\{
      t\left(\bm{k},\bm{k},\omega-\bm{k}\cdot\bm{v}+\frac{\Omega}{2}\right)-
      t^*\left(\bm{k},\bm{k},\omega-\bm{k}\cdot\bm{v}-\frac{\Omega}{2}\right)\right\}g\left(\bm{v}\right)\ud\bm{v}\right]
      f\left(\bm{k},\bm{q},\omega,\Omega\right) =
      2i\Im\left[\bra G\left(\bm{k},\omega\right)\ket\right]
   \\\label{bethe-salpeter2} & \hspace{1cm}
      \times \rho
      \int t\left(\bm{k},\bm{k}',\omega-\bm{k}\cdot\bm{v}+\frac{\Omega}{2}\right)t^*\left(\bm{k},\bm{k}',\omega-\bm{k}\cdot\bm{v}-\frac{\Omega}{2}\right)
      \bm{\delta}\left[\omega'-\omega-\left(\bm{k}'-\bm{k}\right)\cdot\bm{v}\right]
      g\left(\bm{v}\right)f\left(\bm{k}',\bm{q},\omega',\Omega\right)\ud\omega'\ud\bm{v}\frac{\ud^3\bm{k}'}{8\pi^3}.
   \end{align}
\end{widetext}
We can glimpse in this equation the role of each term in the coming transport equation. 
The first two ones correspond respectively to
the temporal and spatial evolution of the specific intensity. The term $t-t^*$ is related to the extinction
coefficient and the integral on the right side of the equation with the term $tt^*$ refers to the scattering process. 

In a dilute system, the imaginary part of the averaged Green function is peaked
and can be written (this is the ``on-shell" approximation)
$ \Im\left[\bra G\left(\bm{k},\omega\right)\ket\right]\approx -\pi\bm{\delta}\left[\omega^2/c_0^2-k^2\right].$
As $\omega$ is closed to $\omega_0$, we can replace $k$ and $k'$ by $k_0$. Finally, defining
the specific intensity $I$ as the Fourier transform of the autocorrelation function of the field (i.e. the so-called
Wigner transform of the field)~\cite{WALTHER-1968}
\begin{equation}\label{specific_intensity}
   f\left(\bm{k},\bm{q},\omega,\Omega\right)=\bm{\delta}\left(k_0-k\right)I\left(\bm{u},\bm{q},\delta,\Omega\right)
\end{equation}
where we recall that $\delta=\omega-\omega_0$, we obtain the Fourier form of the transport equation
\begin{widetext}
   \begin{align}\nonumber
      \left[-\frac{i\Omega}{c_0}+i\bm{u}\cdot\bm{q}+\int_{\bm{v}}\mu_e\left(\delta-k_0\bm{u}\cdot\bm{v},\Omega\right)
      g\left(\bm{v}\right)\ud\bm{v}\right]
      I\left(\bm{u},\bm{q},\delta,\Omega\right)
      = \frac{1}{4\pi}
      \int_{4\pi}\int_{\bm{v}}\int_{-\infty}^{\infty} & \mu_s\left(\delta-k_0\bm{u}\cdot\bm{v},\Omega\right)
      \bm{\delta}\left[\delta'-\delta-k_0\left(\bm{u}'-\bm{u}\right)\cdot\bm{v}\right]
   \\ & \label{transport}
      \times g\left(\bm{v}\right)I\left(\bm{u}',\bm{q},\delta',\Omega\right)\ud\bm{v}\ud\bm{u}'\ud\delta'
   \end{align}
   where
   \begin{align}\label{extinction}
      \mu_e\left(\delta-k_0\bm{u}\cdot\bm{v},\Omega\right) & =\frac{i\rho}{2k_0}\left\{t\left(\omega-k_0\bm{u}\cdot\bm{v}+\frac{\Omega}{2}\right)
      -t^*\left(\omega-k_0\bm{u}\cdot\bm{v}-\frac{\Omega}{2}\right)\right\}
    \\\nonumber
      \textrm{ and }
      \mu_s\left(\delta-k_0\bm{u}\cdot\bm{v},\Omega\right) & =\frac{\rho}{4\pi}t\left(\omega-k_0\bm{u}\cdot\bm{v}+\frac{\Omega}{2}\right)
      t^*\left(\omega-k_0\bm{u}\cdot\bm{v}-\frac{\Omega}{2}\right)
   \end{align}
   where we have written $t\left(\bm{k},\bm{k}',\omega\right)\equiv t\left(\omega\right)$ for the sake of simplicity.
   The coefficients
   $\mu_e$ and $\mu_s$ are like extinction and scattering coefficients
   respectively but depending on time and frequency. 
   The form of the previous equation in real space is
   \begin{align}\nonumber
      & \left[\frac{1}{c_0}\frac{\partial}{\partial t}+\bm{u}\cdot\bm{\nabla}_{\bm{r}}\right]
      I\left(\bm{u},\bm{r},\delta,t\right)
      = -\int_{t'=0}^{\infty}\hat{\mu}_e\left(t',\delta-k_0\bm{u}\cdot\bm{v}\right)g\left(\bm{v}\right)
      I\left(\bm{u},\bm{r},\delta ,t-t'\right)\ud\bm{v}\ud t'
   \\ & \hspace{1cm}\label{transport2}
      +\frac{1}{4\pi}
      \int_{t'=0}^{\infty}\int_{4\pi}\int_{\bm{v}}\int_{-\infty}^{\infty}
      \hat{\mu}\left(t',\delta-k_0\bm{u}\cdot\bm{v}\right)
      \bm{\delta}\left[\delta'-\delta-k_0\left(\bm{u}'-\bm{u}\right)\cdot\bm{v}\right]g\left(\bm{v}\right)
      I\left(\bm{u}',\bm{r},t-t',\delta'\right)\ud\bm{v}\ud\bm{u}'\ud\delta'\ud t'
   \end{align}
\end{widetext}
where $\hat{\mu}_e$ and $\hat{\mu}_s$ are the inverse temporal Fourier transforms of $\mu_e$ and $\mu_s$ respectively.

In the following,
we suppose that
we have a Maxwell-Boltzmann distribution of velocities given by
\begin{equation}\label{distribution}
   g\left(\bm{v}\right) = \frac{1}{\left[\vbar\sqrt{2\pi}\right]^3}\exp\left[-\frac{\bm{v}^2}{2\vbar^2}\right],
\end{equation}
that is a gaussian distribution with vanishing mean and standard deviation $\vbar$.
Eq.~(\ref{transport2}) is the main result of this paper. It is valid whatever $\vbar$. But in the following,
we focus on $k_0\vbar\ll\Gamma$ (Eq.~(\ref{validity-naive})). It has the same structure 
as the well-known radiative transfer equation
(RTE) as derived by Chandrasekhar~\cite{CHANDRASEKHAR-1950} except that it exhibits a frequency coupling and temporal
convolution products.
Note also that the phase function $p\left(\bm{u},\bm{u}'\right)$ (part of an incident beam in the direction $\bm{u}'$
scattered into the direction $\bm{u}$) is constant and equal to $1/4\pi$ because of the
isotropic scattering assumption. As already mentioned, polarization effects can be easily
taken into account by inserting in Eq.~(\ref{transport2}) the full phase function of e.g. a Rayleigh or
a Mie scatterer.

The coupling between the components at various detunings $\delta$ is due to the motion of the scatterers:
the Doppler shift at each scattering event
produces a change of frequency for the scattered photon. 
For a pinned system, $\vbar\to 0,$ the
Gaussian becomes a Dirac-delta function, and the components at various detunings $\delta$
become uncoupled. For each component, one recovers the well-known RTE. 

The convolution
products in time describes the fact that resonances can be responsible 
for a drastic reduction of the velocity of energy propagation in the medium.
This is developed in Sect.~\ref{modal_approach}. Far from resonance, the extinction 
and scattering coefficients do not depend on frequency 
and, again, we recover the well-known RTE.

The optical theorem (or the Ward identity) is verified in Eq.~(\ref{transport2}).
Indeed, we have
\begin{align}\nonumber
      \int_{\bm{v}}
      \mu_e\left(\delta-k_0\bm{u}\cdot\bm{v},\Omega\right)
      = &
      \frac{1}{4\pi}\int_{4\pi}\int_{\bm{v}}\int_{-\infty}^{\infty}
      \mu_s\left(\delta-k_0\bm{u}\cdot\bm{v},\Omega\right)
   \\\nonumber
      \times g\left(\bm{v}\right)\ud{\bm{v}}
      \hspace{1cm}
      &
      \hspace{1cm}\times\bm{\delta}\left[\delta'-\delta-k_0\left(\bm{u}'-\bm{u}\right)\cdot\bm{v}\right]
   \\\nonumber
      &
      \hspace{1cm}\times g\left(\bm{v}\right)\ud\bm{v}\ud\bm{u}'\ud\delta'.
   \\\nonumber
      \underbrace{\hphantom{\int_{\bm{v}}\mu_e\left(\delta-k_0\bm{u}\cdot\bm{v},\Omega\right)=}}_{\textrm{Losses}}
      &
      \underbrace{\hphantom{\frac{1}{4\pi}\int_{4\pi}\int_{\bm{v}}\int_{-\infty}^{\infty}\mu_s\left(\delta-k_0\bm{u}\cdot\bm{v},\Omega\right)}}_{\textrm{Gains}}
\end{align}
No absorption is present in the system: the extinction phenomena is only 
due to scattering. 
More precisely, the photon flux
is conserved but not exactly the energy flux. 
This is because we neglected the recoil effect: when the frequency of the scattered
photon is different from the incoming frequency, the energy difference is in fact
transferred to the atomic scatterer whose velocity is slightly modified.
In our case, the violation of energy conservation is of the order of 
$\Gamma/\omega_0$ and can be safely forgotten.

The steady-state transport
equation for pinned atoms makes it possible to define the scattering mean-free path as
\begin{equation}\label{ell-delta}
   \ell\left(\delta\right)=\lim_{\Omega\to 0}=\frac{1}{\mu_e\left(\delta,\Omega\right)}=
   \ell_0\left[1+\frac{4\delta^2}{\Gamma^2}\right]
\end{equation}
where $\ell_0=k_0^2/\left(4\pi\rho\right)$ is the scattering mean-free path at 
resonance.

\section{Numerical simulations}\label{numerical_sim}

\subsection{A Monte-Carlo scheme}

By integrating over the frequency $\delta'$, the structure of Eq.~(\ref{transport})
is similar to the usual RTE. We have

\begin{widetext}
   \begin{align}\nonumber
      \left[-\frac{i\Omega}{c_0}+i\bm{u}\cdot\bm{q}+\int_{\bm{v}}\mu_e\left(\delta-k_0\bm{u}\cdot\bm{v},\Omega\right)
      g\left(\bm{v}\right)\ud\bm{v}\right]
      I\left(\bm{u},\bm{q},\delta,\Omega\right)
      = & \frac{1}{4\pi}
      \int_{4\pi}\int_{\bm{v}}\mu_s\left(\delta-k_0\bm{u}\cdot\bm{v},\Omega\right)g\left(\bm{v}\right)
   \\\label{transport_final} & \hphantom{\frac{1}{4\pi}\int_{4\pi}\int_{\bm{v}}} \times
      I\left(\bm{u}',\bm{q},\delta+k_0\left(\bm{u}'-\bm{u}\right)\cdot\bm{v},\Omega\right)\ud\bm{v}\ud\bm{u}'.
   \end{align}
\end{widetext}

This specific form explains
why it is possible to solve it using a Monte-Carlo scheme, the only
complication being that we have to deal with complex probability density. 
The details of the computation are given in
appendix~\ref{monte_carlo}. In a word, our original approach consists 
in solving the temporal Fourier transform of the transport
equation and not directly the transport equation itself. This allows us 
to avoid the problem of the temporal convolution products. This
Monte-Carlo scheme is the only one which we could find to work in this case 
and should be used whenever resonances are present in the system. 

\subsection{Slab geometry}

\begin{figure}
   \psfrag{L}{$L$}
   \psfrag{t}{$T_{1/2}$}
   \psfrag{R}{$R\left(t,\delta\right)$}
   \psfrag{T}{$T\left(t,\delta\right)$}
   \includegraphics[width=0.8\linewidth]{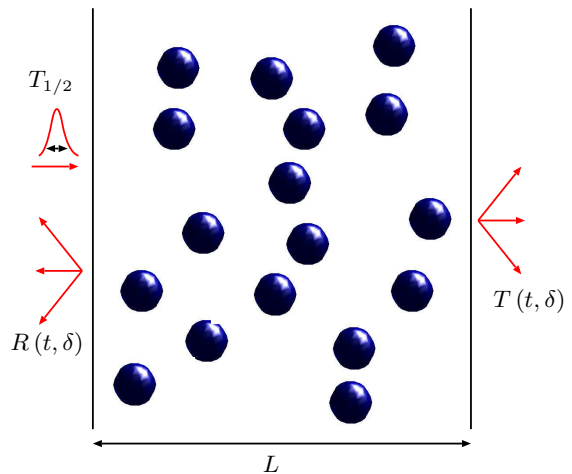}
   \caption{(Color online). Geometry of the system studied numerically. The transmitted
   and reflected fluxes are integrated over all directions.}
   \label{system}
\end{figure}

We  numerically study a cold atomic medium whose shape is a slab, 
infinite along the $x$ and $y$ directions, but extending along 
the $z$ direction in the range $0\leq z\leq L$.
This system is chosen for its simplicity and does not represent a real atomic cloud
released from a standard magneto-optical trap. 
Nevertheless,
this is a true three dimensional geometry.
The slab is illuminated at normal incidence from the left by a gaussian pulse of 
temporal FWHM $T_{1/2}$
at frequency $\omega_L$, detuned from the atomic resonance by
$\delta_L=\omega_L-\omega_0\ll\omega_0$. 
We have $T_{1/2}\gg 2\pi/\omega_L$ so that the
incident pulse is well-defined and its temporal and spectral evolutions 
are independent.
We are interested in the expression of the energy density and the 
reflected and transmitted fluxes denoted
by $U\left(z,t,\delta\right)$, $R\left(t,\delta\right)$ and $T\left(t,\delta\right)$ respectively.
The system and the notations are clarified in Fig.~\ref{system} 
and all the numerical parameters are given in Table~\ref{parameters}.
The computations are done in a cluster of $96$ Intel Xeon
processors at $3.0\,\textrm{GHz}$.

The most important parameter is the ratio of the Doppler shift
to the natural linewidth. As discussed above, we choose it much smaller
than unity:
\begin{equation}
\frac{k_0 \vbar}{\Gamma} = \frac{1}{50}
\end{equation}
The optical thickness at resonance is chosen as:
\begin{equation}
b = \frac{L}{\ell_0} = 10 \ \ \ \mathrm{or} \ \ \ 40
\end{equation}
The first value is chosen so that Eq.~(\ref{validity}) is satisfied so that
the effect of the atomic velocity is expected to be small
and the temporal behavior similar to the one for pinned atoms. In the second case,
important effects are expected.
The atomic density is chosen such that:
\begin{equation}
\rho \lambda_0^3 = 0.1 \ll 1
\end{equation}  
so that the medium is dilute and its index of refraction very close to unity.

This specific choice of parameters for the calculation is such that
the results depend only on the physically
important parameters: $k_0 \vbar/\Gamma,$ $b$ and $\delta/\Gamma.$

\newlength{\thickness}
\setlength{\thickness}{\linewidth}
\divide \thickness by 2
\begin{table}
   \begin{tabular}{@{}p{\thickness}@{}p{\thickness}@{}}
   \hline\hline
      \multicolumn{2}{@{}c@{}}{Parameters and numerical values}
   \\\hline
      $\omega_0/2\pi=3.85\times 10^{14}\,\textrm{Hz}$
   &
      $\Gamma/2\pi=5.9\times 10^{6}\,\textrm{Hz}$
   \\
      $\vbar=9.17\times 10^{-2}\,\textrm{m.s}^{-1}$
   &
      $\rho=2.12\times 10^{17}\,\textrm{m}^{3}$
   \\
      $T_{1/2}=1.00\times 10^{-7}\,\textrm{s}$
   &
      $L=40\ell_0$ or $L=10\ell_0$
   \\\hline\hline 
   \end{tabular}
   \caption{Numerical values of the parameters which are chosen such that 
   all assumptions made to derive the transport
   equation are fulfilled in particular $\rho\lambda_0^3\ll 1$, 
   $k_0\vbar\ll\Gamma$ and $\Gamma\ll\omega_0$. These parameters correspond
   to the Rubidium atom.}
   \label{parameters}
\end{table}

\subsection{Numerical results}

\subsubsection{Quadratic regime}

\begin{figure}
   \psfrag{o}{$\delta/\Gamma$}
   \psfrag{t1}{$\Gamma t=10$}
   \psfrag{t2}{$\Gamma t=20$}
   \psfrag{t3}{$\Gamma t=40$}
   \psfrag{t4}{$\Gamma t=80$}
   \psfrag{u}{Relative transmission}
   \includegraphics[width=\linewidth]{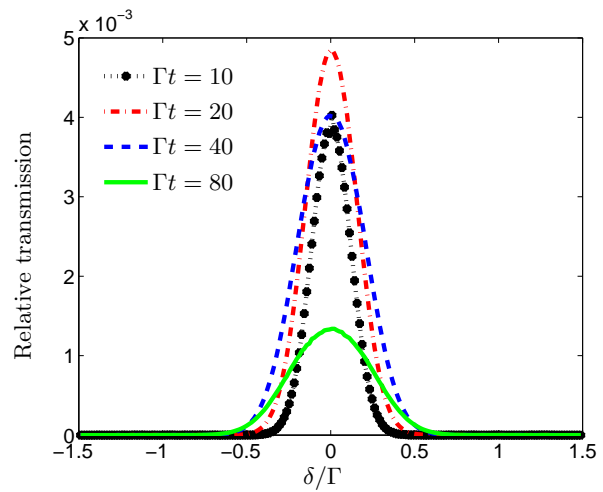}
   \caption{(Color online). Spectral distribution of the transmitted photons for five different times in the case of a thin
   slab of optical thickness $b=10.$ The spectral broadening is due to the accumulation along a multiple
   scattering path of small Doppler
   shifts (here typically 0.02 $\Gamma$ at each scattering event).
   At short time, the broadening is small compared to the natural linewidth $\Gamma,$ resulting in a Gaussian
   lineshape. At long times, 
   the spectral distribution broadening saturates, with an approximate Gaussian lineshape.
   Note that the total transmitted flux is larger at $\Gamma t=20$ than at $\Gamma t=10,$ because
   the photons has to be multiply scattered before escaping in the forward direction. The decay
   at longer time is due to the finite trapping time in the medium. 
   }
   \label{transmitted_thin}
\end{figure}

\begin{figure*}
   \psfrag{t}{$\Gamma t$}
   \psfrag{t1}{$\Gamma t=160$}
   \psfrag{t2}{$\Gamma t=320$}
   \psfrag{t3}{$\Gamma t=490$}
   \psfrag{t4}{$\Gamma t=810$}
   \psfrag{o}{$\delta/\Gamma$}
   \psfrag{s}{$\vbar=0$}
   \psfrag{o1}{$\delta/\Gamma=0$}
   \psfrag{o2}{$\delta/\Gamma=0.40$}
   \psfrag{o3}{$\delta/\Gamma=0.80$}
   \psfrag{u}{Relative transmission}
   \psfrag{v}{Relative transmission}
   \includegraphics[width=\linewidth]{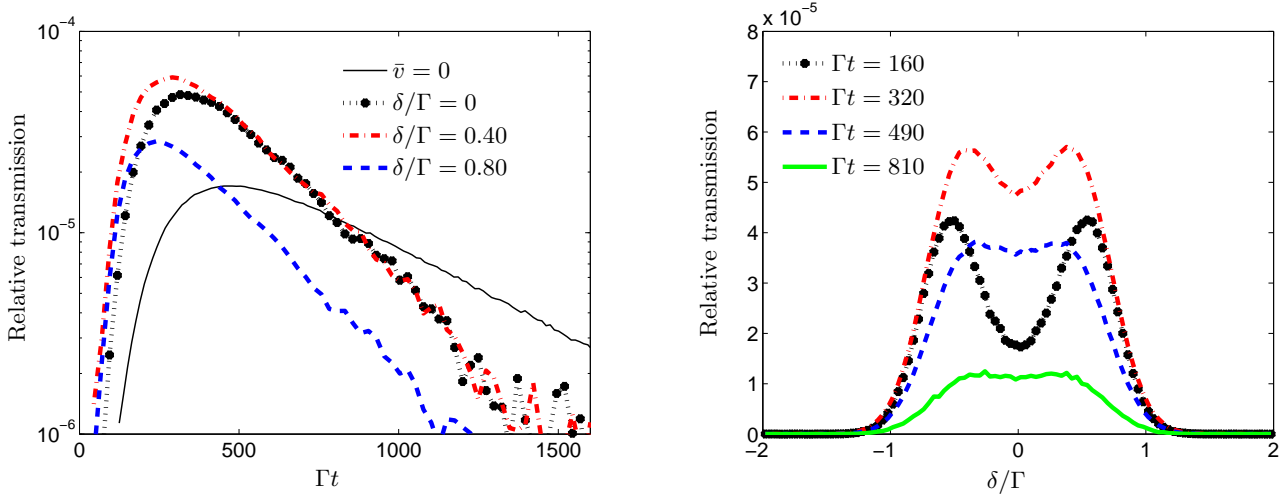}
   \caption{(Color online). Transmission $T\left(t,\delta\right)$ versus time for three different 
   frequencies (left plot) and 
   transmission versus frequency for four different times (right plot). We have also plotted the temporal
   behavior of the transmission for a pinned system (i.e. $\vbar=0$). The incoming pulse
   is at resonance $\delta_L=0$, the optical thickness of the medium is $b=40$ and $k\vbar/\Gamma=0.02$.
   Note that the time scale which governs the temporal evolution
   is large compared to $\Gamma^{-1}$ while the spectral scale is of the order of $\Gamma^{-1}$.
   This validates the independence of the temporal and spectral variables in the specific intensity.}
   \label{transmitted}
\end{figure*}

When a system with sufficiently small optical thickness $b=10$ is illuminated at resonance (i.e. $\delta_L=0$),
we expect the photon to escape before the frequency broadening induced by Doppler
effect becomes important. 
The result is shown in Fig.~\ref{transmitted_thin}.
At short time, one observes
distinctly a Gaussian lineshape with linewidth $\ll \Gamma.$ This
is the regime where the photon frequency performs a random walk, and thus
a global diffusive behaviour is seen in frequency space.
Most photons escape the medium before their frequency is significantly
shifted. 
At longer time, the frequency dependence
of the scattering mean-free path cannot be ignored. The width of the spectral distribution 
tends to saturate and the lineshape flattens at the center.
 This is the quadratic regime where the decay rate
$\tau$ is quadratic in $L$ as for pinned atoms~\cite{DELANDE-2003}.

\subsubsection{Doppler regime}

We now study a larger system --- optical thickness $b=40$ --- for
which the Doppler shift is important. 
The numerical results for the transmitted flux at various frequencies are given in Fig.~\ref{transmitted}.
The temporal behavior is an exponential decay at long times. Importantly,
the decay rate is \emph{identical} $\tau=262\Gamma^{-1}$ for all frequency components, although each
component has a different scattering mean free path in the medium and consequently
a different optical thickness (it varies by a factor 3.5 between $\delta=0$ and $\delta=0.8 \Gamma.$)
However, the time for this regime to settle depends on the frequency. More precisely, 
since many scattering events are needed to create photons in the system with large detunings, 
this time increases with $\delta$.
This regime is studied
in details in Sect.~\ref{modal_approach}. The spectral behavior exhibits two peaks which is very surprising 
at first glance.
This comes from the interplay between the variation of the scattering mean-free path with the frequency 
(characterized by $\Gamma$)
and the frequency redistribution due to the Doppler shift (characterized by the width of the Gaussian shape $\vbar$). 
More precisely, since the mean free-path increases with the detuning, for example, $\ell(\delta=\Gamma)=5\ell_0$, a resonant photon
is quite efficiently trapped, whereas a far detuned photon escapes rapidly. Therefore, resonant
photons undergo a lot of Doppler shifts, populating the other frequencies. On the contrary, far detuned photons
most likely escape without any frequency shift. At a certain frequency, these two effects are balancing each other,
giving rise to a maximum photon population, hence, a two peak structure in the transmission.
Note that the symmetry observed in the spectrum
of the transmitted flux is due to the fact that the illumination is at $\delta_L=0$ and all the 
properties of the system are even functions of $\delta$. This would be no longer exactly true if
recoil effects were taken into account.

\begin{figure*}
   \psfrag{t}{$\Gamma t$}
   \psfrag{t1}{$\Gamma t=160$}
   \psfrag{t2}{$\Gamma t=320$}
   \psfrag{t3}{$\Gamma t=490$}
   \psfrag{t4}{$\Gamma t=810$}
   \psfrag{o}{$\delta/\Gamma$}
   \psfrag{s}{$\vbar=0$}
   \psfrag{o1}{$\delta/\Gamma=0$}
   \psfrag{o2}{$\delta/\Gamma=0.40$}
   \psfrag{o3}{$\delta/\Gamma=0.80$}
   \psfrag{u}{Density of photons (arbitrary units)}
   \psfrag{v}{Density of photons (arbitrary units)}
   \includegraphics[width=\linewidth]{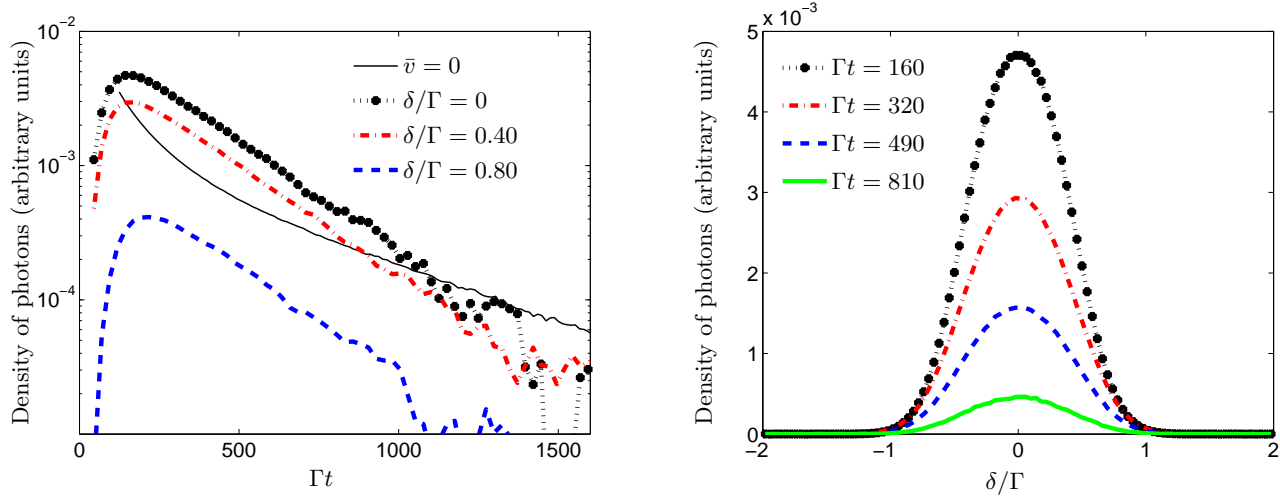}
   \caption{(Color online). Density of photons $U\left(L/2,t,\delta\right)$ at the center of the slab versus 
   time for three different frequencies (left
   plot) and energy density versus frequency for fiour different times (right plot). We have also plot the temporal
   behavior of the transmission for a pinned system (i.e. $\vbar=0$).
   The optical thickness of the system is $b=40$.}
   \label{energy_density}
\end{figure*}

The time dependence of the energy density at the center of the slab (i.e. for $z=L/2$) plotted in
Fig.~\ref{energy_density} is similar to the one of the transmitted
flux. In particular, the exponential decay at long times is the same. Concerning the spectral dependence, we also
see two peaks which are nevertheless attenuated at very long times. The reason is the same as for the transmitted flux:
trapping of photons is more efficient at resonance. The spectrum at long times is also studied in detail
in Sect.~\ref{modal_approach} using a modal approach.

If we continue to increase the optical thickness of the system, the exponential decay becomes lower
(i.e. $\tau\to+\infty$ when $L\to+\infty$) and the spacing between the peaks increases. In any case, the 
decay rate is very different from the one $\tau\propto (L/\ell_0)^2\Gamma^{-1}$ for pinned atoms and also
different from the one deduced from the Holstein equation (i.e. 
$\tau\propto (L/\ell_0)\sqrt{\log\left[L/(2\ell_0)\right]}\Gamma^{-1}$~\cite{HOLSTEIN-1947,MOLISCH-1998}). We call this  the
Doppler regime.

If the temperature of the cold atomic gas is increased (at fixed optical thickness),
the temporal decay becomes faster (i.e. $\tau\to 0$ when $\vbar\to+\infty$) and the spacing between the two peaks
in the spectral distribution 
increases.

\begin{figure}
   \psfrag{o}{$\delta/\Gamma$}
   \psfrag{t1}{$\Gamma t=40$}
   \psfrag{t2}{$\Gamma t=120$}
   \psfrag{t3}{$\Gamma t=240$}
   \psfrag{t4}{$\Gamma t=810$}
   \psfrag{u}{Relative transmission}
   \includegraphics[width=\linewidth]{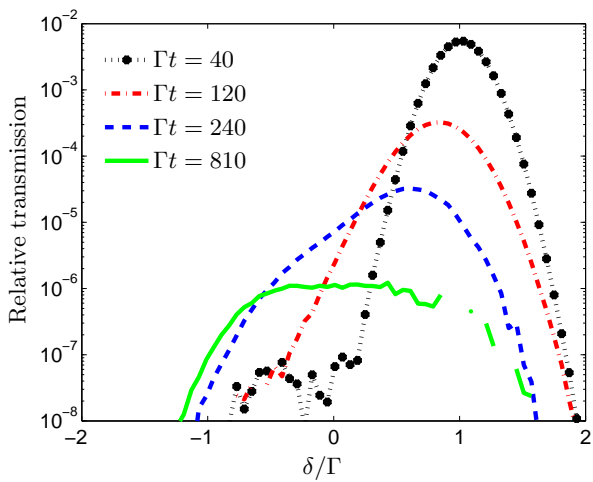}
   \caption{(Color online). Relative transmission $T\left(t,\delta\right)$ versus frequency for four different times 
   in the case of a detuned excitation $\delta_L=\Gamma.$
   The spectral behavior at long time tends to be symmetric, as in the resonant case. 
   The optical thickness of the system at resonance is 40.}
   \label{transmitted_detuned}
\end{figure}

Figure~\ref{transmitted_detuned} gives an example of calculation for a detuned excitation at $\delta_L=\Gamma$. 
One observes at short time a broadening of the spectral distribution. Soon, this broadening
becomes asymmetric, photons being ``attracted" towards resonance, simply because there
are more efficiently trapped near resonance. In addition, the height of the peak on the other side
of the resonance increases. Finally, at long times, 
the spectral shape becomes identical to the one of Fig.~\ref{energy_density}, independently of the initial detuning:
there, multiple Doppler shifts have erased the memory of the initial spectral distribution.

\section{Temporal and spectral behavior at long times}\label{modal_approach}

Analysing analytically the temporal and spectral shape at long times and for large systems usually ends up with
the derivation of a diffusion equation which describes the evolution of the energy density.
The main advantage of a transport equation
compared to a diffusion equation is that it contains all transport regimes of photons from ballistic
to diffusive~\cite{PIERRAT-2006}. Despite these drawbacks, the diffusion approximation is widely used, in particular
in biological imaging~\cite{ISHIMARU-1997,YODH-1995,GAYEN-1996} because of its simplicity.
In this section, we first derive the diffusion approximation from
the transport equation for the cases of pinned and moving atoms.
To achieve this goal, we use a modal decomposition of the specific intensity to obtain a dispersion
relation which leads to the expression of the diffusion coefficient.

\subsection{Fokker-Planck equation}

As mentioned above, the residual Doppler broadening is usually much smaller 
(by about two orders of magnitude)
than the natural width $\Gamma$ in the MOT regime, so that Eq.~(\ref{validity-naive}) is valid. 
Thus, we can perform a second order Taylor expansion in $k\vbar/\Gamma$
of the extinction and scattering coefficients, and of the specific intensity.
We cannot restrict the expansion to the first order because of the symmetry of $g\left(\bm{v}\right)$.
This gives
\begin{align}\nonumber
   \mu_{e,s}\left(\delta-k_0\bm{u}\cdot\bm{v},\Omega\right) \sim &
   \mu_{e,s}\left(\delta,\Omega\right)
   -k_0\bm{u}\cdot\bm{v}\mu_{e,s}'\left(\delta,\Omega\right)
\\\nonumber & 
   + \frac{k_0^2\left(\bm{u}\cdot\bm{v}\right)^2}{2}\mu_{e,s}''\left(\delta,\Omega\right)
\\\nonumber
   I\left(\bm{u},\bm{q},\delta-k_0\bm{u}\cdot\bm{v},\Omega\right) \sim &
   I\left(\bm{u},\bm{q},\delta,\Omega\right)
   -k_0\bm{u}\cdot\bm{v}I'\left(\bm{u},\bm{q},\delta,\Omega\right)
\\\nonumber & 
   + \frac{k_0^2\left(\bm{u}\cdot\bm{v}\right)^2}{2}I''\left(\bm{u},\bm{q},\delta,\Omega\right)
\end{align}
where the symbols $'$ and $''$ denote $\partial/\partial\delta$ and $\partial^2/\partial\delta^2$ respectively.
By integrating over the velocity of the atoms, the transport equation (i.e. Eq.~(\ref{transport_final}))
becomes a Fokker-Planck type equation which reads in the Fourier space
\begin{widetext}
   \begin{align}\nonumber
      & \left[-\frac{i\Omega}{c_0}+i\bm{u}\cdot\bm{q}+\mu_e\left(\delta,\Omega\right)
      +\frac{\left(k_0\vbar\right)^2}{2}\mu_e''\left(\delta,\Omega\right)\right]
      I\left(\bm{u},\bm{q},\delta,\Omega\right)
      =\frac{1}{4\pi}
      \int_{4\pi}\left[\vphantom{\frac{\left(k_0\vbar\right)^2}{2}}
         \mu_s\left(\delta,\Omega\right)I\left(\bm{u}',\bm{q},\delta,\Omega\right)
         -\bm{u}\cdot\left(\bm{u}'-\bm{u}\right)\left(k_0\vbar\right)^2
      \right.
   \\\label{fokker_planck} & \quad\quad\left.
         \times\mu_s'\left(\delta,\Omega\right)I'\left(\bm{u}',\bm{q},\delta,\Omega\right)
         +\frac{\left(k_0\vbar\right)^2}{2}\mu_s''\left(\delta,\Omega\right)I\left(\bm{u}',\bm{q},\delta,\Omega\right)
         +\frac{\left\|\bm{u}-\bm{u}'\right\|^2\left(k_0\vbar\right)^2}{2}\mu_s\left(\delta,\Omega\right)I''\left(\bm{u}',\bm{q},\delta,\Omega\right)
      \right]\ud\bm{u}'.
   \end{align}
\end{widetext}
For the sake of simplicity, we consider the slab geometry mentioned in Sect.~\ref{numerical_sim}. 
This choice does not reduce the generality of the derivation, the system being a true three dimensional
geometry. This permits a simplification of the Fokker-Planck equation given in Eq.~(\ref{fokker_planck})
by integrating over $x$ and $y$ (invariance of the problem by translation along $x$ and $y$)
and $\varphi$ (invariance of the problem by rotation around $z$) where $\theta$ and $\varphi$ are
the usual spherical angles. To obtain an analytical derivation, the terms $\bm{u}\cdot\left(\bm{u}'-\bm{u}\right)$
and $\left\|\bm{u}-\bm{u}'\right\|$ have to be approximated. We denote by
$\cos\Theta=\bm{u}\cdot\bm{u}'$ the cosine of the scattering angle. So we have
\begin{equation}\nonumber
   \bm{u}\cdot\left(\bm{u}'-\bm{u}\right) = \cos\Theta-1
   \textrm{ and }\left\|\bm{u}-\bm{u}'\right\| =\sqrt{2\left(1-\cos\Theta\right)}.
\end{equation}
For large systems and at long times, many scattering events occur. So we can replace the cosine of the
scattering angle by its average given by the so-called anisotropy factor. In our case, the scattering being
isotropic (the phase function is constant), the anisotropy factor vanishes. Thus we write
$\bm{u}\cdot\left(\bm{u}'-\bm{u}\right)\sim-1$ and
$\left\|\bm{u}-\bm{u}'\right\|\sim\sqrt{2}$ and the Fokker-Planck equation reduces to
\begin{widetext}
   \begin{align}\nonumber
      & \left[-\frac{i\Omega}{c_0}+i\mu q+\mu_e\left(\delta,\Omega\right)
      +\frac{\left(k_0\vbar\right)^2}{2}\mu_e''\left(\delta,\Omega\right)\right]
      I\left(\mu,q,\delta,\Omega\right)
      =\frac{1}{2}
      \int_{-1}^{+1}\left[\vphantom{\frac{\left(k_0\vbar\right)^2}{2}}
         \mu_s\left(\delta,\Omega\right)I\left(\mu',q,\delta,\Omega\right)
         +\left(k_0\vbar\right)^2
         \mu_s'\left(\delta,\Omega\right)I'\left(\mu',q,\delta,\Omega\right)
      \right.
   \\\label{fokker_planck2} & \quad\quad\quad\quad\left.
         +\frac{\left(k_0\vbar\right)^2}{2}\mu_s''\left(\delta,\Omega\right)I\left(\mu',q,\delta,\Omega\right)
         +\left(k_0\vbar\right)^2\mu_s\left(\delta,\Omega\right)I''\left(\mu',q,\delta,\Omega\right)
      \right]\ud\mu'.
   \end{align}
\end{widetext}
where $\mu=\cos\theta$ is the direction.
This equation can be simplified assuming that we are looking for large systems such that $L\gg \ell_0$ 
(i.e. $q\ll \pi/\ell_0$ and $\Omega\ll\Gamma$). 
At detuning $\delta$, the propagating time between two scattering events is of the order
of $\tau_{\textrm{prop}}=\ell\left(\delta\right)/c_0$. On the other hand, the Wigner time delay 
characteristic of the scattering process is about $\tau_{\textrm{sca}}=\Gamma^{-1}$. Then we can assume
that $\tau_{\textrm{prop}}\ll\tau_{\textrm{sca}}$ and consider that the speed of light in vacuum
is infinite. In other words, the propagation of energy in the system is hugely
reduced by radiation trapping --- an effect enhanced by the resonant character of the
interaction with the atomic scatterers --- so that free propagation in the vacuum
can be considered as instantaneous.
Using these approximations, Eq.~(\ref{fokker_planck2}) rewrites
\begin{widetext}
   \begin{equation}\label{fokker_planck3}
      \left[i\mu q+\left(1-i\frac{\Omega}{\Gamma}\right)
      \beta\left(\delta\right)\right]
      I\left(\mu,q,\delta,\Omega\right)
      =\frac{1}{2}
      \int_{-1}^{+1}\left[
         \beta\left(\delta\right)
         I\left(\mu',q,\delta,\Omega\right)
         +\gamma\left(\delta\right)I'\left(\mu',q,\delta,\Omega\right)
         +\eta\left(\delta\right)I''\left(\mu',q,\delta,\Omega\right)
      \right]\ud\mu'
   \end{equation}
   where we have defined
   \begin{align}\nonumber
      \beta\left(\delta\right) & = \frac{1}{\ell\left(\delta\right)}+\frac{\left(k_0\vbar\right)^2}{2}\left(\frac{1}{\ell\left(\delta\right)}\right)''
      =\frac{1}{\ell_0\left(1+4\delta^2/\Gamma^2\right)}\left[1-4\left(\frac{k_0\vbar}{\Gamma}\right)^2\frac{1-12\delta^2/\Gamma^2}{\left(1+4\delta^2/\Gamma^2\right)^2}\right],
   \\\nonumber
      \gamma\left(\delta\right) & = \left(k_0\vbar\right)^2\left(\frac{1}{\ell\left(\delta\right)}\right)'
      =-8\left(\frac{k_0\vbar}{\Gamma}\right)^2\frac{\delta}{\ell_0\left(1+4\delta^2/\Gamma^2\right)^2},
   \\\nonumber
      \eta\left(\delta\right) & = \frac{\left(k_0\vbar\right)^2}{\ell\left(\delta\right)}
      =\frac{\left(k_0\vbar\right)^2}{\ell_0\left(1+4\delta^2/\Gamma^2\right)}.
   \end{align}
   Eq~(\ref{fokker_planck3}) reads in the real space
   \begin{align}\nonumber
      \left[\mu\frac{\partial}{\partial z}+\beta\left(\delta\right)\frac{1}{\Gamma}\frac{\partial}{\partial t}\right]
      I\left(\mu,z,\delta,t\right)
      = & -\beta\left(\delta\right)I\left(\mu,z,\delta,t\right)
   \\\label{fokker_planck3_bis} & 
      +\frac{1}{2}
      \int_{-1}^{+1}\left[
         \beta\left(\delta\right)
         I\left(\mu',z,\delta,t\right)
         +\gamma\left(\delta\right)\frac{\partial}{\partial\delta}I\left(\mu',z,\delta,t\right)
         +\eta\left(\delta\right)\frac{\partial^2}{\partial\delta^2}I\left(\mu',z,\delta,t\right)
      \right]\ud\mu'.
   \end{align}
\end{widetext}
The first two terms on the left hand side of Eq.~(\ref{fokker_planck3_bis}) represent the spatio-temporal
evolution of the specific intensity. The first term on the right hand side describes losses due to scattering.
The integral term over the direction and the second derivative in frequency represent gains by scattering in the real
and frequency spaces respectively.
Finally the first derivative in frequency is a consequence of a drift in the Doppler process
which tends to shift the frequency towards the resonance.

\subsection{Modal approach}

We now expand the specific intensity in the form
\begin{align}\label{modes}
   I\left(z,\mu,t,\delta\right)=
   \int_{-\infty}^{+\infty}
   \sum_{s\left(q\right)}g_{q,s}\left(\mu,\delta\right)\exp\left[iq z+s\left(q\right)t\right]\frac{\ud q}{2\pi}.
\end{align}
To obtain this expression, we have performed a spatial Fourier transform of the specific intensity. 
For each $q$,
there exist several eigenvalues $s\left(q\right),$ which may be complex or real, whose spectrum may
be discrete or continuous. $g_{q,s}$ is the associated
eigenvector which depends on direction and frequency only. This decomposition is a generalized form
of the one used to derive the diffusion approximation starting from the standard RTE~\cite{PIERRAT-2006}.
For a system of size $L$, the dominant $q$ is given by $q=\pi/\left(L+2z_0\right),$ where 
$z_0$ is the so-called extrapolation length~\cite{ROSSUM-1999}, of the order
of the mean-free path, accounting for the effects at the interface between the medium and the vacuum. 
For an optically thick medium, $ b \gg 1,$ one has $z_0\ll L$.
Therefore, the mode surviving at large time, i.e. having the longest lifetime in the system, corresponds to the eigenvalue $s_0\left(\pi/L\right)$ with the lowest (in magnitude) negative real part. 
Inserting Eq.~(\ref{modes}) into Eq.~(\ref{fokker_planck3}) leads to
\begin{align}\nonumber
   & \left[i\mu q+\left(1+\frac{s}{\Gamma}\right)
   \beta\left(\delta\right)\right]
   g_{q,s}\left(\mu,\delta\right)
   =\frac{1}{2}
   \int_{-1}^{+1}\left[
      \beta\left(\delta\right)
      g_{q,s}\left(\mu',\delta\right)
   \right.
\\\label{fokker_planck4} & \quad\left.
      +\gamma\left(\delta\right)g_{q,s}'\left(\mu',\delta\right)
      +\eta\left(\delta\right)g_{q,s}''\left(\mu',\delta\right)
   \right]\ud\mu'.
\end{align}
Thus the eigenvector is given by
\begin{align}\nonumber
   & g_{q,s}\left(\mu,\delta\right)=
   \frac{1}{2\left[iq\mu+\left(1+s/\Gamma\right)\beta\left(\delta\right)\right]}
   \int_{-1}^{+1}\left[
      \beta\left(\delta\right)g_{q,s}\left(\mu',\delta\right)
   \right.
\\\nonumber & \hspace{1cm}\left.
      +\gamma\left(\delta\right)g_{q,s}'\left(\mu',\delta\right)
      +\eta\left(\delta\right)g_{q,s}''\left(\mu',\delta\right)
   \right]\ud\mu'.
\end{align}
By integrating over the direction, $\mu$, we obtain
\begin{align}\nonumber
   g_{q,s}\left(\delta\right)= & \frac{1}{q}
   \arctan\left[
      \frac{q}{\left(1+s/\Gamma\right)\beta\left(\delta\right)}
   \right]
   \left[
      \beta\left(\delta\right)g_{q,s}\left(\delta\right)
   \right.
\\\label{dispersion} & \left.
      +\gamma\left(\delta\right)g_{q,s}'\left(\delta\right)
      +\eta\left(\delta\right)g_{q,s}''\left(\delta\right)
   \right].
\end{align}
This linear second order differential equation can be seen as a generalized Schr\"odinger problem. 
Note that this is not exactly an eigenvalue
problem since the eigenvalue is not in front of the associated eigenvector.

\subsection{Pinned atoms}

For pinned atoms, $\vbar$ is equal to zero. Eq.~(\ref{dispersion}) decouples
in an independent equation for each frequency $\delta,$ with the following dispersion relation for the lowest eigenvalue $s_0$:
\begin{equation}\nonumber
   \frac{1}{q\ell\left(\delta\right)}
   \arctan\left[\frac{q\ell\left(\delta\right)}{1+s_0/\Gamma}\right]=1
\end{equation}
In an experiment, the relevant $\delta$ value is fixed by the frequency of the excitation $\delta_L$. 
For sufficiently large systems --- larger than the mean-free path at frequency $\delta$, see Eq.~(\ref{ell-delta}) ---
we have $q\ell\left(\delta\right) \ll 1$
which corresponds to the diffusion approximation regime. A Taylor expansion in the dispersion relation leads to
\begin{equation}\nonumber
   s_0=-\frac{q^2\ell^2\left(\delta\right)\Gamma}{3}.
\end{equation}
The $q^2,$ i.e. $1/L^2$, dependence is a characteristic of the quadratic diffusion regime~\cite{DELANDE-2003}.
This leads to the definition of the diffusion coefficient $\mathcal{D}$
and the velocity of energy transport $c_{\textrm{tr}}$ by
\begin{equation}\nonumber
   \mathcal{D}=-\frac{s_0}{q^2}=\frac{c_{\textrm{tr}}\left(\delta\right)\ell\left(\delta\right)}{3}
\end{equation}
with 
\begin{equation}\label{c-transport}
c_{\textrm{tr}}=\ell\left(\delta\right)\Gamma = c_0 \ \frac{\Gamma}{\omega_0}\ \frac{k_0^3}{4\pi\rho}\ \left(1+ \frac{4\delta^2}{\Gamma^2}\right).
\end{equation}
The first expression of $c_{\textrm{tr}}$ has a simple physical interpretation: a single scattering event followed
by propagation on a mean-free path takes time $\Gamma^{-1}.$ Note that this is true whatever the detuning is, a highly non-trivial
property: it turns out that both the Wigner time delay at scattering and the time delay during propagation (at the
group velocity) depend on the detuning, but not their sum.  This property has been used in ~\cite{LABEYRIE-2005} for 
simple Monte-Carlo simulations. The present, more rigorous approach, justifies these simple simulations.

The second expression for $c_{\textrm{tr}}$ is simply the product of the vacuum light velocity $c_0$ by the inverse
of the quality factor of the atomic resonance, by (up to a numerical factor) the
number of atoms per cubic wavelength, and finally by a factor being unity at resonance.
Because the quality factor $\omega_0/\Gamma$ of an atomic resonance can be very high, typically $10^8,$ the velocity
can be strongly diminished even in a dilute medium where $\rho/k_0^3 \ll 1.$ 
In typical experimental situations, it can be reduced by 4 orders of magnitude~\cite{LABEYRIE-2005}.
This is the main characteristic of radiation trapping by resonant scatterers.
These results are in perfect agreement with 
literature~\cite{TIP-1991,TIP-1992,OZRIN-1992,LAGENDIJK-1993,MINIATURA-2002}.

The temporal exponential decay rate at long times is then given by
\begin{equation}\nonumber
   \tau=-\frac{1}{s_0}=\frac{3L^2}{\pi^2\ell\left(\delta\right)^2\Gamma},
\end{equation}
a result also in complete agreement with literature~\cite{MOLISCH-1998,LABEYRIE-2005}.
In a word, defining a diffusion coefficient for fixed scatterers is possible and is even well-understood. The atomic
nature of the scatterers induces a strong decrease of the velocity of energy propagation inside the system.
This property could be used to design new high capacity quantum memories.

\subsection{Moving atoms}

In the case of moving atoms, the detuning $\delta$ may range from $-\infty$ to $+\infty.$, 
such that the inequality $q\ell_0\delta^2/\Gamma^2\ll 1$ cannot hold for all $\delta$, 
ruling out an approach based on a Taylor expansion. 
Physically, it means that for large detunings, the mean-free path is larger than the
system size, invalidating the diffusion approximation.
As a consequence, it is not possible to define
a global diffusion constant in such a system and the evolution of the energy
density cannot be described with a diffusion equation. The same situation occurs for the CFR case and
was underlined in~\cite{KENTY-1932,MOLISCH-1998}. This explains in particular the necessity
to write an integro-differential transport equation~\cite{HOLSTEIN-1947}.

\begin{figure}
   \psfrag{L}{$L/\ell_0$}
   \psfrag{t}{$\tau\Gamma$}
   \psfrag{p1}{Full Monte Carlo simulation}
   \psfrag{p2}{Dispersion relation}
   \psfrag{p3}{Quadratic behavior}
   \psfrag{p4}{Holstein prediction}
   \includegraphics[width=\linewidth]{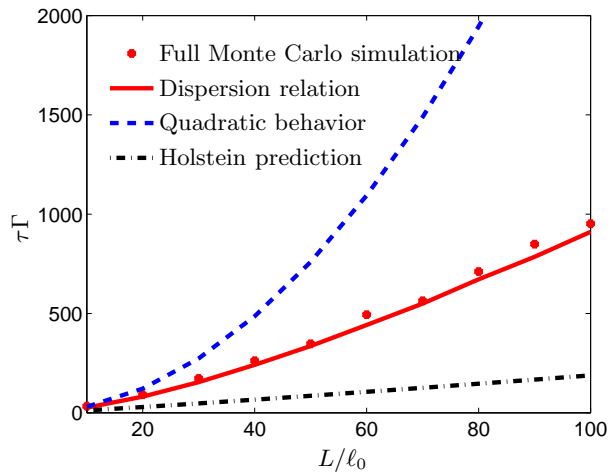}
   \caption{(Color online). Temporal decay rate $\tau$ of the outgoing flux and the energy density
   versus the system size. Crosses correspond to the full Monte-Carlo simulations (i.e. resolution
   of Eq.~(\ref{transport_final})), the solid line is related to the numerical resolution of the dispersion relation
   (Eq.~(\ref{dispersion})). For small $L$ (but large enough to be in the diffusive regime), the behavior
   is quadratic as for pinned scatterers (dotted line). This is the quadratic regime, which extends
   roughly up to $L/\ell_0\approx 30,$ as predicted by Eq.~(\ref{def-calA}). For very large systems,
   the behavior is completely different from the Holstein
   prediction (dashed line). This is the Doppler regime. The agreement between the two numerical methods is very good,
   the small difference coming from approximating the cosine of the scattering angle by its average (anisotropy factor).}
   \label{tau}
\end{figure}

To the best of our knowledge, it is not possible
to obtain an analytical form of the eigenvalues $s$ from Eq.~(\ref{dispersion}).
In order to make some comparisons with the results
given by the numerical resolution of Eq.~(\ref{transport_final}), we have numerically solved Eq.~(\ref{dispersion})
by a straightforward shooting technique. 
The results are shown in Fig.~\ref{tau}. The agreement is excellent,
only small deviations
being visible for extremely large systems. Note that the results deviate very significantly from
the quadratic behaviour and from the Holstein prediction using the CFR assumption.
This proves that resonant cold atomic media are fundamentally different from usual hot gases as far as radiation
trapping is concerned.

\subsubsection{``Diffusion equation''}

For large
systems such that the optical thickness at resonance is large, $q\ell_0 \ll 1$ and if we look
at times longer than $\Gamma^{-1}$, we can use directly on Eq.~(\ref{transport2}) the same approximation
used for deriving Eq.~(\ref{fokker_planck3}) to perform the integration over $t'.$
We obtain the following simplified transport equation:
\begin{align}\nonumber
   & \left[
      \bm{u}\cdot\bm{\nabla}_{\bm{r}}+\left(1+\frac{1}{\Gamma}\frac{\partial}{\partial t}\right)
      \int_{\bm{v}}\frac{g\left(\bm{v}\right)\ud\bm{v}}{\ell\left(\delta-k_0\bm{u}\cdot\bm{v}\right)}
   \right]
   I\left(\bm{u},\bm{r},\delta,t\right)
\\\nonumber
   & = \frac{1}{4\pi}\int_{4\pi}\int_{\bm{v}}
   \frac{g\left(\bm{v}\right)}{\ell\left(\delta-k_0\bm{u}\cdot\bm{v}\right)}
   I\left(\bm{u}',\bm{r},\delta+k_0\left(\bm{u}'-\bm{u}\right)\cdot\bm{v},t\right)
\\\label{transport5}
   & \hphantom{= \frac{1}{4\pi}\int_{4\pi}\int_{\bm{v}}}
   \times\ud\bm{v}\ud\bm{u}'
\end{align}
Thus this transport equation is what we can call the ``diffusion equation'' for the system at finite temperature.
A simple physical
picture can then be extracted from this approximation in term of a Monte-Carlo scheme. Note that Eq.~(\ref{transport5})
is valid whatever $\vbar$. It
is exactly the same as the standard time-dependent RTE except that the Doppler broadening is present and the
energy velocity depends on frequency. The Monte-Carlo method is then the same. When a photon enters the system
at the frequency $\delta_L$, it propagates over a distance whose average is
$1/\left[\int_{\bm{v}}g\left(\bm{v}\right)\ud\bm{v}/\ell\left(\delta_L-k_0\bm{u}\cdot\bm{v}\right)\right]$
(this is the first term on the right hand side).
Then it is scattered isotropically (the last term with the double integral)
and undergoes a Doppler drift (from $\delta+k_0\left(\bm{u}'-\bm{u}\right)\cdot\bm{v}$ to $\delta)$. 
Next, it propagates again with a new scattering
mean-free path. Each scattering process takes about $\Gamma^{-1}$. The implementation of this method is
easier than the one detailed in Sect.~\ref{monte_carlo} and gives reliable results at large scales only.

This simple picture can be used to find a quantitative criterion to discriminate between the Doppler and the quadratic regimes too.
As explained in the introduction, at large $b$, a typical multiple scattering path can be viewed as a random walk of the photon
in the medium with step $\ell$ approximately. Thus the photon is multiply scattered about $b^2$ times before
escaping. The Doppler shift is on average a random variable with zero average and 
about $(k\vbar)^2$ variance. If successive scattering events are statistically independent,
the photon frequency performs itself a random walk with step about $k\vbar,$
and the typical accumulated Doppler shift after $N$ scattering events is of the order
of $\sqrt{N}k\vbar.$ As $N\sim b^2=L^2/\ell^2_0,$  the key parameter is thus
\begin{equation}\label{def-calA}
   \mathcal{A}=\frac{k_0\vbar}{\Gamma}\frac{L}{\ell_0}.   
\end{equation}
If $\mathcal{A}\ll 1$, we are in the quadratic regime and the temporal decay rate $\tau$ is given by the diffusion
approximation prediction (as for fixed atoms). This is nothing but the inequality~(\ref{validity}) derived in the 
introduction using hand waving arguments.
For $\mathcal{A}\gg 1$, we are in the Doppler regime where
the expression of the temporal decay rate is no longer quadratic in $L.$

\subsubsection{Energy density and outgoing fluxes}

Knowing the expression of the eigenvalue, it is possible to obtain analytically the behavior of the eigenvector close 
to resonance (i.e. for $\delta\to 0$).
By using a Taylor expansion of $g_{q,s}$ at second order in Eq.~(\ref{dispersion}), 
we obtain the second derivative of the eigenvector at the resonant frequency
given by
\begin{equation}\nonumber
   g_{\pi/L,s_0}''\left(\delta\right)\underset{\delta\to 0}{\sim}\frac{1}{k_0^2\vbar^2}\left[\frac{s_0}{\Gamma}+\frac{\pi^2\ell_0^2}{3L^2}\right].
\end{equation}

Using the previous results on the modal approach for the specific intensity for large systems at long times,
we can easily deduce the expressions of the energy density and the outgoing fluxes close to resonance. Keeping only the
lowest mode, the specific intensity writes
\begin{equation}\label{fundamental_mode}
   I\left(z,\mu,t,\delta\right)=g_{\pi/L,s_0}\left(\mu,\delta\right)
   \sin\left[\frac{\pi}{L}z\right]\exp\left[s_0 t\right].
\end{equation}
Thus, the energy density is given by
\begin{equation}\nonumber
   U\left(z,t,\delta\right)=\frac{g_{\pi/L,s_0}}{c_{\textrm{tr}}\left(\delta\right)}
   \left(\delta\right)\sin\left[\frac{\pi}{L}z\right]\exp\left[s_0 t\right].
\end{equation}
Using the expression of the energy velocity, Eq.~(\ref{c-transport}), and the expression of the
second derivative of the eigenvector near resonance, we obtain the spectrum of the energy density
near the resonance in the form
\begin{equation}\nonumber
   U\left(\delta\right)\underset{\delta\to 0}{\propto}
   1+\left\{
      \frac{1}{2k_0^2\vbar^2}\left[\frac{s_0}{\Gamma}+\frac{\pi^2\ell_0^2}{3L^2}\right]-\frac{4}{\Gamma^2}
   \right\}\delta^2.
\end{equation}
This expression shows that it is impossible to have two peaks in the spectrum of the energy density for large
systems at long times, because the second derivative remains negative whatever the size of the system.
This is in complete agreement with the full Monte Carlo numerical simulations. To derive
the same result for the outgoing fluxes, we have to integrate over the directional and the space variables the
Fokker-Planck equation (i.e. Eq.~(\ref{fokker_planck3_bis})) which gives
\begin{align}\nonumber
   & \int_{-1}^{+1}\mu\left[I\left(\mu,L,\delta,t\right)-I\left(\mu,0,\delta,t\right)\right]\ud\mu
   =\left(1+\frac{1}{\Gamma}\frac{\partial}{\partial t}\right)
   \beta\left(\delta\right)
\\\nonumber &
   \int_{-1}^{+1}\int_0^L I\left(\mu,z,\delta,t\right)\ud z\ud\mu
   +
   \int_{-1}^{+1}\int_0^L
   \left[
      \beta\left(\delta\right)
   \right.
\\\nonumber & \quad
      \times I\left(\mu',z,\delta,t\right)
      +\gamma\left(\delta\right)I'\left(\mu',z,\delta,t\right)
\\\label{transport6} & \quad\left.
      +\eta\left(\delta\right)I''\left(\mu',z,\delta,t\right)
   \right]\ud z\ud\mu'.
\end{align}
At long times, the entering flux vanishes. So the first term of Eq.~(\ref{transport6})
is the total outgoing flux denoted by $\phi\left(t,\delta\right)$.
This is the sum of the reflected ($R\left(t,\delta\right)$) and transmitted ($T\left(t,\delta\right)$) fluxes.
Actually, at long times, a dynamical equilibrium is reached inside the system and the spectra of the reflected
and transmitted fluxes are identical. Inserting the expression of the specific intensity reduced to the
lowest mode, Eq.~(\ref{fundamental_mode}), and using the dispersion relation, Eq.~(\ref{dispersion}),
the total outgoing flux writes
\begin{align}\nonumber
   \phi\left(t,\delta\right)
   = & 2\left\{
      \left[\arctan\left(\displaystyle\frac{\pi}{L\left(1+s_0/\Gamma\right)\beta\left(\delta\right)}\right)\right]^{-1}
   \right.
\\ & \left.\hphantom{2\{}
      -\frac{L\left(1+s_0/\Gamma\right)\beta\left(\delta\right)}{\pi}
   \right\}g_{\pi/L,s_0}\left(\delta\right)\exp\left(s_0t\right).
\end{align}
Using the same result as for the energy density, we obtain the second order approximation in frequency of the outgoing flux as
\begin{align}\nonumber
   \phi\left(\delta\right)\underset{\delta\to 0}{\propto}
   1+\left\{
      \frac{1}{2k_0^2\vbar^2}\left[\frac{s_0}{\Gamma}+\frac{\pi^2\ell_0^2}{3L^2}\right]+\frac{4}{\Gamma^2}
   \right\}\delta^2.
\end{align}
The sign of the second derivative of the flux may change and become positive for very large systems (i.e. $L\to\infty$ and $s_0\to 0$)
which is at the root of the observation of two peaks on the reflected and transmitted fluxes.
This is also in complete agreement with the numerical results.
The physical picture is quite clear: far detuned photons are relatively rare in the medium, because
they are less trapped and escape more rapidly: this is why they manifest themselves in the transmitted and
reflected fluxes more strongly than in the energy density in the bulk.

\section{Conclusion}\label{conclusion}

In this paper, we have derived a transport equation for the incoherent radiation propagating in a
cold atomic gas at finite temperature. The derivation is based on first principles generalized to the case of moving scatterers.
This equation, valid for the case of partial frequency redistribution, is fully justified by microscopic
arguments. It is solved numerically by an original and fully justified Monte-Carlo scheme which gives 
reliable results on the outgoing
fluxes and the energy density for all optical thicknesses and incident beams. 
A modal approach is used to obtain information
on the spectral and temporal behaviors at long times. The main result is that the temporal behavior
is a single exponential decay for all frequency components and the outgoing
fluxes can exhibit two spectral peaks for very large systems.
The velocity of energy propagation in the system is also affected by the motion of the atoms but is still reduced
by typically four orders of magnitude compared to its velocity in vacuum. This is an important result
if we think of quantum memory applications for such systems.

We would like to thank the French Centre National de la Recherche Scientifique (CNRS) for financial support.
CPU time was provided by the Institut Francilien de Recherche sur les Atomes Froids (IFRAF).

\appendix

\section{Spatio-temporal Fourier transforms of relevant quantities}\label{fourier_transforms}

Here is the list of all conventions used for the spatio-temporal Fourier transforms of the Green functions $G_0$, $\bra G\ket$, $G$
(respectively in the vacuum, in the scattering system on average or not, all denoted by $G$ in the following), of the mass 
operator $M_m$, of the intensity operator $K_m$ and of the scattering operator $t_m^{\bm{v}}$.
   \begin{align}\nonumber
      G\left(\bm{r},t\right)= &
      \int G\left(\bm{k},\omega\right)\exp\left[i\bm{k}\cdot\bm{r}-i\omega t\right]\frac{\ud^3\bm{k}}{8\pi^3}\frac{\ud\omega}{2\pi},
   \\\nonumber
      M_m\left(\bm{r},t\right)= &
      \int M_m\left(\bm{k},\omega\right)\exp\left[i\bm{k}\cdot\bm{r}-i\omega t\right]\frac{\ud^3\bm{k}}{8\pi^3}\frac{\ud\omega}{2\pi},
   \end{align}
   \begin{align}\nonumber
      & K_m\left(\bm{r}_1,\bm{r}_3,\bm{r}_2,\bm{r}_4,t_1,t_3,t_2,t_4\right)=
   \\\nonumber & \hspace{1cm}
      \int K_m\left(\bm{k}_1,\bm{k}_3,\bm{k}_2,\bm{k}_4,\omega_1,\omega_3,\omega_2,\omega_4\right)
   \\\nonumber & \hspace{1cm}\hphantom{\int}\times
      \exp\left[i\bm{k}_1\cdot\bm{r}_1-i\bm{k}_3\cdot\bm{r}_3-i\bm{k}_2\cdot\bm{r}_2+i\bm{k}_4\cdot\bm{r}_4\right]
   \\\nonumber & \hspace{1cm}\hphantom{\int}\times
      \exp\left[-i\omega_1 t_1+i\omega_3 t_3+i\omega_2 t_2-i\omega_4 t_4\right]
   \\\nonumber & \hspace{1cm}\hphantom{\int}\times
      \frac{\ud^3\bm{k}_1}{8\pi^3}\frac{\ud^3\bm{k}_2}{8\pi^3}\frac{\ud^3\bm{k}_3}{8\pi^3}\frac{\ud^3\bm{k}_4}{8\pi^3}
      \frac{\ud\omega_1}{2\pi}\frac{\ud\omega_2}{2\pi}\frac{\ud\omega_3}{2\pi}\frac{\ud\omega_4}{2\pi},
   \end{align}
   \begin{align}\nonumber
      & t_m\left(\bm{r}_1,\bm{r}_2,t_1,t_2\right)=
      \int t_m\left(\bm{k}_1,\bm{k}_2,\omega_1,\omega_2\right)
   \\\nonumber & \hspace{1cm}\times
      \exp\left[i\bm{k}_1\cdot\bm{r}_1-i\bm{k}_2\cdot\bm{r}_2-i\omega_1 t_1+i\omega_2 t_2\right]
   \\\nonumber & \hspace{1cm}\times
      \frac{\ud^3\bm{k}_1}{8\pi^3}\frac{\ud^3\bm{k}_2}{8\pi^3}
      \frac{\ud\omega_1}{2\pi}\frac{\ud\omega_2}{2\pi}.
   \end{align}
   The signs in the exponentials are chosen such that the vertex operator $K_m$ describes correctly the correlation between the following two scattering processes:
   \begin{equation}
      \bm{k}_3,\omega_3\rightarrow\bm{k}_1,\omega_1\quad\bm{k}_4,\omega_4\rightarrow\bm{k}_2,\omega_2
   \end{equation}
   where $\bm{k}_3,\bm{k}_4$ and $\bm{k}_1,\bm{k}_2$ are the incident and emergent wave-vectors respectively. 
   $\omega_3,\omega_4$ and $\omega_1,\omega_2$ are the incident and emergent frequencies respectively.

\section{Monte Carlo scheme}\label{monte_carlo}

The temporal dependence of Eq.~(\ref{transport2}) does not allow us to use a Monte-Carlo scheme to solve it numerically.
Nevertheless, its spatio-temporal Fourier transform as written in Eq.~(\ref{transport_final}) has the same structure as the
standard RTE in the steady-state regime. This is the reason why we are using a Monte-Carlo type simulation. In practice
the implementation is the same as for the standard RTE~\cite{HAMMERSLEY-1964,FISHMAN-1996} except that the probability
densities are different and a Fourier transform is needed. Defining an effective extinction coefficient by
\begin{equation}\nonumber
   \mu_{e,\textrm{eff}}\left(\delta,\Omega\right)=\int_{\bm{v}}\mu_e\left(\delta-k_0\bm{u}\cdot\bm{v},\Omega\right)g\left(\bm{v}\right)\ud\bm{v}
   -\frac{i\Omega}{c_0},
\end{equation}
this equation writes
\begin{align}\nonumber
   & I\left(\bm{q},\bm{u},\Omega,\delta\right) =
   \frac{1}{4\pi}\int_{4\pi}\int_{\bm{v}}\mu_s'\left(\delta-k_0\bm{u}\cdot\bm{v},\Omega\right)
\\\label{transport3}
   & \times g\left(\bm{v}\right)
   \frac{I\left(\bm{q},\bm{u}',\Omega,\delta+k_0\left(\bm{u}'-\bm{u}\right)\cdot\bm{v}\right)}{i\bm{u}\cdot\bm{q}+\mu_{e,\textrm{eff}}\left(\delta,\Omega\right)}
   \ud\bm{v}\ud\bm{u}'.
\end{align}
To go back to the real space for the space variable, we remark that $\Re\left[\mu_{e,\textrm{eff}}\left(\delta,\Omega\right)\right]$
is strictly positive for all $\Omega$ and $\delta$ which implies that
\begin{equation}
   \frac{1}{i\bm{q}\cdot\bm{u}+\mu_{e,\textrm{eff}}}=\int_0^{\infty}\exp\left[-\left(i\bm{q}\cdot\bm{u}+\mu_{e,\textrm{eff}}\right)s\right]\ud s
\end{equation}
and then
\begin{equation}
   \operatorname{FT}_{\bm{q}}\left[\frac{1}{i\bm{q}\cdot\bm{u}+\mu_{e,\textrm{eff}}}\right]
   =\int_0^{\infty}\bm{\delta}\left(\bm{r}-\bm{u}s\right)\exp\left[-\mu_{e,\textrm{eff}} s\right]\ud s
\end{equation}
where $\operatorname{FT}_{\bm{q}}$ denotes the Fourier transform operator over $\bm{q}$. Finally, the integral
form of the transport equation writes
\begin{align}\nonumber
   & I\left(\bm{r},\bm{u},\Omega,\delta\right)=
   \frac{1}{\mu_{e,\textrm{eff}}\left(\delta,\Omega\right)}\int_{s=0}^{+\infty} \mu_{e,\textrm{eff}}\left(\delta,\Omega\right)\exp\left[-\mu_{e,\textrm{eff}}\left(\delta,\Omega\right)s\right]
\\\nonumber
   &\hspace{0.5cm} \times\frac{1}{4\pi}\int_{4\pi}\int_{\bm{v}}
   \mu_s\left(\delta-k_0\bm{u}\cdot\bm{v},\Omega\right)
\\
   &\hspace{0.5cm} \times I\left(\bm{r}-s\bm{u},\bm{u}',\Omega,\delta+k_0\left(\bm{u}'-\bm{u}\right)\cdot\bm{v}\right)\ud\bm{v}\ud\bm{u}'\ud s.
\end{align}
This form fully justifies the use of a Monte-Carlo scheme. First, we have to compute the extinction coefficient for all frequencies $\Omega$
and detunings $\delta$. This coefficient is given by Eq.~(\ref{extinction}) (Voigt profile):
\begin{align}\nonumber
   \mu_{e,\textrm{eff}}\left(\delta,\Omega\right)= &\frac{4i\pi\rho}{2k_0^2}\int \left[\frac{\Gamma/2}{\delta-k_0\bm{u}\cdot\bm{v}+\Omega/2+i\Gamma/2}\right.
   \\ & \nonumber
                                                           \left.        -\frac{\Gamma/2}{\delta-k_0\bm{u}\cdot\bm{v}-\Omega/2-i\Gamma/2}
         \right]
   \\ & \nonumber
   \times\frac{1}{\left[\vbar\sqrt{2\pi}\right]^3}\exp\left[-\frac{\bm{v}^2}{2\vbar^2}\right]\ud\bm{v}
        -\frac{i\Omega}{c_0}.
\end{align}
To compute it, we use a Gauss-Hermite quadrature. Then, we perform the Monte-Carlo simulation
The probability densities used to compute the integrals are the following:
\begin{itemize}
   \item the probability density to have an extinction process (in the sense of the coefficient $\mu_{e,\textrm{eff}}$)
   at the position $s$ without having one from $0$ to $s$ is $p_s\left(s\right)=\mu_{e,\textrm{eff}}\left(\delta,\Omega\right)\exp\left[-\mu_{e,\textrm{eff}}\left(\delta,\Omega\right)s\right]$;
   \item the probability density for a photon to be scattered in the direction $\bm{u}$ coming from the direction $\bm{u}'$
   is $p_{\bm{u}}\left(\bm{u},\bm{u}'\right)=1/\left(4\pi\right)$;
   \item the probability density for an atom to have the velocity $\bm{v}$ is given by $g\left(\bm{v}\right)$. Thus, the new frequency
   is given by $\delta=\delta'-k_0\left(\bm{u}'-\bm{u}\right)\cdot\bm{v}$.
\end{itemize}
The main difficulty is that the probability density $p_s$ is complex. In that case, we have to deal with the modulus of $p_s$
as follows
\begin{equation}
   \int p_s\left(s\right)f\left(s\right)\ud s=\lim_{n\to\infty}\frac{1}{n}\sum_{j=1}^n
   \frac{p_s\left(s_j\right)}{\left|p_s\left(s_j\right)\right|}f\left(s_j\right).
\end{equation}
where $f:\mathbb{R}\rightarrow\mathbb{C}$. The elements $s_j$ are distributed in respect of the probability density $\left|p_s\right|$.
Note that the error bars on $I\left(\bm{r},\bm{u},\Omega,\delta\right)$ in the Monte-Carlo simulation have to be sufficiently
small for the Fourier transform to give reliable results on $I\left(\bm{r},\bm{u},t,\delta\right)$.


\end{document}